\documentclass[12pt,preprint]{aastex}

\newcommand{\msol}{M$_{\odot}$}

\newcommand{\escs}{erg\,s$^{-1}$cm$^{-2}$sr$^{-1}$}
\newcommand{\cmt}{cm$^{-3}$}
\newcommand{\cmd}{cm$^{-2}$}

\newcommand{\kms}{\,km\,s$^{-1}$}
\newcommand{\um}{$\,\mu$m}

\newcommand{\wat}{H$_2$O}
\newcommand{\Ox}{[OI]}
\newcommand{\Oxline}{[OI]\,63$\,\mu$m}

\shorttitle{\Oxline\, jets in class 0 sources}
\shortauthors{Nisini et al.}

\begin{document}


\title{\Oxline\, jets in class 0 sources detected by \textit{Herschel}}


%
%
%


\author{B. Nisini\altaffilmark{1}, G. Santangelo\altaffilmark{1,2}, T. Giannini\altaffilmark{1}, S. Antoniucci\altaffilmark{1}, 
S. Cabrit\altaffilmark{3}, C. Codella\altaffilmark{2}, C.J. Davis\altaffilmark{4}, J. Eisl\"offel\altaffilmark{5}, L. Kristensen\altaffilmark{6}
G. Herczeg\altaffilmark{7}, D. Neufeld\altaffilmark{8}, E.F. van Dishoeck\altaffilmark{9,10} }

\altaffiltext{1}{INAF -- Osservatorio Astronomico di Roma, via Frascati 33, I-00040 Monte Porzio, Italy}
\altaffiltext{2}{INAF -- Osservatorio Astrofisico di Arcetri, Largo E. Fermi 5, 50125 Firenze, Italy}
\altaffiltext{3}{LERMA, Observatoire de Paris, UMR 8112 of the CNRS, 61 Av. de L'Observatoire, 75014 Paris, France}
\altaffiltext{4}{Astrophysics Research Institute, Liverpool John Moores University, Liverpool, UK}
\altaffiltext{5}{Th\"uringer Landessternwarte, Sternwarte 5, 07778, Tautenburg, Germany}
\altaffiltext{6}{Harvard-Smithsonian Center for Astrophysics, 60 Garden Street, Cambridge, MA, 02138, USA}
\altaffiltext{7}{Kavli Institute for Astronomy and Astrophysics, Peking University, Yi He Yuan Lu 5, Haidian Qu, Beijing 100871, China}
\altaffiltext{8}{Department of Physics and Astronomy, Johns Hopkins University, 3400 North Charles Street, Baltimore, MD 21218, USA}
\altaffiltext{9}{Sterrewacht Leiden, Leiden University, PO Box 9513, 2300 RA, Leiden, The Netherlands}
\altaffiltext{10}{Max-Planck-Institut für extraterrestriche Physik, Postfach 1312, 85741, Garching, Germany}

\begin{abstract}

We present \textit{Herschel} PACS mapping observations of the \Oxline\, line towards protostellar outflows 
in the L1448, NGC1333-IRAS4, HH46, BHR71 and VLA1623 star forming regions. 
We detect emission spatially resolved along the outflow direction,
which can be associated with a low excitation atomic jet. In the L1448-C, 
HH46~IRS and BHR71~IRS1 outflows this
emission is kinematically resolved into blue- and red-shifted jet lobes, having radial velocities up to 200 \kms.  
In the L1448-C atomic jet the velocity
increases with the distance from the protostar, similarly to what observed in the  SiO jet associated with this source. This suggests that \Ox\, and molecular
gas are kinematically connected and that this latter could represent the colder cocoon of a jet at higher excitation. 
Mass flux rates (\.M$_{jet}$(OI)) have been measured from the \Oxline\, luminosity
adopting two independent methods. We find values in the range 1-4$\times$10$^{-7}$ \msol\,yr$^{-1}$ for all sources
but HH46, for which an order of magnitude higher value is estimated. 
\.M$_{jet}$(OI) are compared with mass accretion rates (\.M$_{acc}$) onto the protostar and with \.M$_{jet}$ derived from ground-based CO observations. 
\.M$_{jet}$(OI)/\.M$_{acc}$ ratios are in the range 0.05-0.5, similar to the values for more evolved sources. 
\.M$_{jet}$(OI) in HH46~IRS and IRAS4A are comparable to \.M$_{jet}$(CO), while those of the remaining sources are significantly  
lower than the corresponding \.M$_{jet}$(CO). 
We speculate that for these three sources most of the mass flux is carried out by a molecular jet, while the warm atomic gas does not significantly contribute to the dynamics of the system.

\end{abstract}





\section{Introduction}

Collimated and high velocity mass ejections are often associated with Young Stellar Objects (YSOs), 
being a natural outcome of the protostellar accretion process.
Indeed, collimated jets have been observed in a variety of YSOs of different evolutionary stages, 
from the embedded class 0 protostars to pre-main sequence Classical T Tauri stars (CTTs), although
the excitation conditions, and consequently the suitable observational tracers, vary
significantly for the different sources (e.g. Frank et al. 2014). 
Jets from CTTs have been investigated in the most detail, since they are optically visible and can 
be studied through several emission lines tracer of hot gas
($\ga$ 10$^4$~K, e.g. Hartigan et al. 2007; Podio et al. 2012).
More embedded class I sources are mainly observed in the near-IR. Compared to CTT jets, class I jets are denser, colder and
less ionized, showing also a significant molecular component (e.g. Davis et al. 2011, Garcia Lopez et al. 2008).

In the less evolved and heavily obscured class 0 objects jets are only sporadically 
detected in the sub-mm as collimated high velocity emission of CO and SiO gas at relatively low excitation
($\sim$ 100-300 K, e.g. Gueth \& Guilloteau 1999; Tafalla et al. 2004; Codella et al. 2007; Hirano et al. 2010). 
Most of the class 0 sources, however, are associated with large scale low velocity molecular CO emission,
tracing cavities or swept-up gas, without any clear evidence of the underlying jet or wind 
pushing away the ambient gas. The above scenario poses the following questions: what is the agent driving the large scale 
molecular outflows in class 0 sources? Is the low excitation collimated molecular jet seen in some class 0 sources  
dynamically important in the mass accretion/ejection balance? Why do only a few class 0 sources have 
collimated molecular jets?

Indeed, the study of jets in class 0 sources suffers from a strong observational bias. 
Since these sources are still deeply embedded in their parental cloud, any optical/near-IR
emission near the source escapes from detection. In addition, given the expected dense environment, it is likely that a significant 
part of the jet material is composed of neutral atomic gas at relatively low temperatures (e.g.
$\sim$ 1000-4000 K) whose emission lines fall in the mid- and far-IR spectral range. 
Gas with such properties has been revealed by recent \textit{Spitzer} observations of young stars, 
showing the presence of low excitation forbidden lines of [SiII], [SI] and [FeII]
close to the central protostar (Dionatos et al. 2009, Neufeld et al. 2009, Lahuis et al. 2010, Baldovin-Saavedra et 
al. 2011).

Determining the dynamical properties of this atomic jet is extremely important in order to
understand the feedback of the active accretion on the surrounding medium and the formation of large scale 
CO outflows. In addition, the mass flux rate carried out by such a jet can
give information on the ejection/accretion efficiency and, when compared to sources with different
ages, its evolution. 

Among the tracers of low excitation atomic jets, the [OI] 63\um\, line is expected to be 
the brightest, due to the high abundance of atomic oxygen and favorable excitation conditions. In particular, 
for pre-shock densities $<$ 10$^5$ \cmt\, the \Oxline\, 
line is predicted to be the main coolant in the original wind shock, i.e. the 
dissociative shock caused by the impact of the high velocity collimated wind ejected
from the protostar on the surrounding material (Hollenbach \& McKee 1989). 
As such, its luminosity is directly  proportional to the wind mass flux rate for a wide 
range of shock conditions (Hollenbach 1985). 

Strong \Ox\, emission has been observed in the past towards young protostars and their associated
outflows, with the first detections obtained by the \textit{Kuiper Airborne Observatory} (KAO, Ceccarelli et al. 1997) and subsequent systematic surveys performed with the \textit{Infrared Space Observatory} (\textit{ISO}, 
Giannini et al. 2001, Nisini et al. 2002, Liseau et al. 2006). In these studies, the importance
of \Oxline\, line emission on the protostellar gas cooling has been addressed, together with its role 
as a tracer of the original wind driving molecular outflows (Giannini et al. 2001).
However, the coarse spatial and spectral resolution of these \textit{ISO} observations (80\arcsec and $R \sim$ 300) 
yields limited morphological and kinematical information on the \Ox\, emission which does not
allow us to unequivocally associate it to the protostellar jet. 
The PACS (Photodetector Array Camera and Spectrometer, Poglitsch et al. 2010) instrument on-board the 
\textit{Herschel space observatory} (Pilbratt et al. 2010) provides a factor of ten improvement
in spatial resolution with respect to \textit{ISO}, and spectral resolution R$\sim$3500, 
allowing one to detect weak lines on a strong continuum and resolve gas at velocities $\ga$ 100 \kms
(e.g. Van Kempen et al. 2010, Karska et al. 2013). 
In addition, the mapping capability of PACS, although limited,  is well suited to 
cover in few observational settings not only the protostellar environment but also 
most of the outflow region. 

PACS observations of the \Oxline\, line have been performed on samples of YSOs of both 
class 0/I (Karska et al. 2013, Green et al. 2013) and CTT (Podio et al. 2012) type, as well as towards individual 
shock regions along outflows (Benedettini et al. 2013, Santangelo et al. 2013). 
These observations were obtained with a single footprint measurement on the targeted objects, 
thus providing only limited spatial information. 
In these studies, the \Oxline\, emission was compared with other tracers of gas
cooling, namely H$_2$O, CO and OH, as well as with available shock and PDR models.
The results of this comparison  all concur to indicate that \Ox\, emission both
close to the source and on specific shock positions comes from high velocity dissociative
shocks, and not from the shock responsible for the molecular emission, confirming
previous findings obtained with \textit{ISO}.

Here we present extensive  PACS mapping of the \Ox\, lines at 63 and 145\um\, in a sample of 
outflows from young low mass protostars, located in the  L1448, NGC1333-IRAS4, BHR71, VLA1623 and HH46
star forming regions.  The sources have been chosen among outflows observed by \textit{ISO} and/or \textit{Spitzer} for 
being associated with
warm gas emission and having extended outflows.

In particular, this paper is focused on the identification of the atomic jet emission associated 
with the large scale molecular
outflows and at inferring its role in the dynamics of the system. 
In Section 2, we describe the observations performed, discuss our data reduction procedure,
and present the \Oxline\, line maps. The morphology of the observed emission
is described in Section 3, where we also discuss the kinematics of the 
\Oxline\, gas in the sources where the line has been resolved in velocity (namely L1448-C, HH46~IRS and 
BHR71~IRS1). In Section 4 we present the methods adopted to derive the mass flux rate carried out
by the atomic gas traced by the \Oxline\, emission, discussing the associated uncertainty 
and assumptions. Finally, in Section 5 we discuss how the measured \Oxline\, mass flux rate 
can be related to the molecular outflow thrust and to the protostellar mass accretion rate.

\section{Observations}

Observations with the PACS instrument in the L1448, NGC1333-IRAS4, BHR71 and VLA1623 regions were 
performed as part of the OT1 program \textit{Probing the physics and dynamics of the hidden warm gas 
in the youngest protostellar outflows} (OT1$\_$bnisini$\_$1). 
In addition, a map of the \Oxline\, emission along the HH46 IRS outflow, obtained as part of the \textit{WISH 
(Water In Star formation with Herschel)} key program (van Dishoeck et al., 2011), 
is also presented here. Table 1 provides the central coordinates of the maps together with 
distance of the region and luminosity of the outflow driving source considered in the analysis.

PACS Integral Field Unit (IFU) line spectroscopy
was used in chopping/nodding mode to obtain spectral maps of the outflows 
at the \Ox\, transition $^3P_1 - ^3P_2$ (63.1852\um\, observed in the second grating order). For the four outflows
of the OT1$\_$bnisini$\_$1 program, simultaneous observations of the $^3P_0 - ^3P_1$ (145.5254\um)
transition (in the first PACS grating order) were also taken.

The IFU consists of a 5$\times$5 spatial pixel (called spaxels) array
providing a spatial sampling of 9$\farcs$4/spaxel, for a total field of view of 
47$\arcsec \times 47\arcsec$. The diffraction-limited FWHM beam size at 63\um\,
(145\um\,) is $\sim$5\arcsec\, (11\arcsec).

The outflows were covered through Nyquist-sampled raster maps arranged
along the outflow axis. The extent of the mapped regions is delineated in Figs. 1-5 as green polygons.
The nominal spectral resolutions of the grating at 63\um\ and 145\um\ are $R\sim$3500 and 1200,
respectively, which correspond to the velocity resolutions of $\sim$ 86 and 250 \kms, respectively.
In flight calibrations have however shown that the PACS effective spectral resolution at 63\um\ could span from 100 to 130\kms\,
(according to the PACS observer manual\footnote{http://herschel.esac.esa.int/Docs/PACS/html/pacs\_om.html}). 
To evaluate the instrumental FWHM in our observations, we have extracted
a \Oxline\, spectrum towards the PDR region located NE of VLA1623 (see Appendix A), under the assumption 
that the intrinsic line-width is negligible (i.e. below 10 \kms). The linewidth of 
the unresolved line is on the order of 130\kms, which we hereafter consider 
our instrumental limit. 

The data of the L1448, NGC1333-IRAS4, BHR71 and VLA1623 sources were acquired 
with a sampling of 44\kms\, per spectral element (i.e. Nyquist sampling at the nominal resolution).  
A factor of two higher spectral sampling was instead adopted for
the observations of HH46 (22\kms\, per spectral element).
The observations were performed with a single scan cycle, providing an integration 
time per spectral resolution element of 30 sec.

All the data were reduced with HIPE\footnote{HIPE is a joint development 
by the \textit{Herschel} Science Ground Segment Consortium, consisting of ESA, 
the NASA \textit{Herschel} Science Center, and the HIFI, PACS and
SPIRE consortia} v10.0, where they were flat-fielded and 
flux-calibrated by comparison with observations of Neptune. The flux calibration uncertainty 
amounts to around 20$\%$ while the \textit{Herschel} pointing accuracy is  
$\sim$ 2$\arcsec$. Finally, in-house IDL routines were
used to locally fit and remove the continuum emission and to construct integrated 
line maps. 

\section{\Ox\, morphology and kinematics}

Figures 1-5 show the \Oxline\, maps of the 5 five observed outflows.
The morphology of the \Ox\ 145\um\, line emission is very similar
to that of the 63\um\ line but images have a lower S/N and will not be presented here. 
The 145\um\, data will be used to estimate the gas density, as described in section 4.1.1.. 
The main outflow properties and the general distribution of the \Oxline\, emission are discussed in Appendix A. 
The obtained maps cover several protostars, most of them associated with their own outflow.
Here we discuss in more detail only the properties of the most prominent outflows in each
region, and discuss the morphology and kinematics of the gas close to the central source, 
where we expect the emission to be connected with the presence of the atomic jet. 

From Figs 1-5, we see that in all sources the \Ox\, emission is spatially resolved close to the source 
and elongated in the direction of the outflow axis. The FWHM width in the direction perpendicular to the 
outflow is instead consistent with the width of the PSF at 63\um\, and thus it
appears non resolved transversely. We therefore interpret the observed emission as coming from
a collimated jet and/or from a series of shocked knots aligned along the outflow axis.

For each source, a spectrum from a rectangular region  encompassing the \Ox\, jet emission 
has been extracted. Table 2 summarizes the spatial extent of these regions together with
the parameters measured for the \Ox\, line emission. The LSR line velocity 
 and width at FWHM have been measured by fitting the spatially
integrated \Oxline\, profile with a Gaussian function. Due to the factor $\sim$ 3 lower 
spectral resolution at 145\um\, this line is always unresolved, so we have not extracted any velocity
information from its profile. 

The observed peak velocities in the integrated spectra are not centered at systemic velocity but
span between -53\kms\, (in BHR71) and +56\kms\, (in IRAS4A), i.e. values comparable
with the spectral sampling. Spurious wavelength shifts 
can also be caused by a misalignment of a point-like emitting source with respect to the spaxel
center\footnote{see sect. 4.7.2 of the PACS observer manual at http://herschel.esac.esa.int/Docs/PACS/html/pacs\_om.html}. 
Although the observed emission is extended over several spaxels,
and thus the wavelength shifts should be minimized, we conservatively assume that 
actual V$_{LSR}$ of the \Oxline , averaged in the extracted spectra, is between the observed
 $V_{LSR}$ and the systemic velocity. 

In L1448-C, BHR71~IRS1 and HH46~IRS the 63\um\, line width is significantly larger than the 
PACS instrumental profile, indicating that the line is spectrally resolved with
an intrinsic width (measured as $\Delta V_{line} = \sqrt{\Delta V_{obs}^2 - \Delta V_{instr}^2}$) on
the order of 110, 90 and 130 \kms\, for L1448, BHR71 and HH46, respectively. 
On the other hand, VLA1623 and IRAS4A have line widths consistent with the instrumental 
profile and thus are not spectrally resolved. 

An enlargement of the region close to the driving source is also presented in Figs 1-5. Here,
the \Ox\, emission, shown as contour plots,  is superimposed on archival \textit{Spitzer}
IRAC images and, when available, on \wat\, 179\um\, Herschel/PACS maps
obtained together with the \Ox\, observations (Bjerkeli et al., 2012, Nisini et al. 2013, Santangelo et al., 2014,
Nisini et al. in prep.). 
For L1448-C, BHR71 and HH46 IRS we have
separately integrated the blue- and red-shifted line profiles to reveal 
spatial variations in the gas kinematics. In these sources, the blue and red outflow
lobes are spatially separated and are in agreement with the kinematics of the outflow
as observed through mm tracers (e.g. Bachiller et al. 1990, Bourke et al. 1997, van Kempen et al. 2009). 

In L1448-C and VLA1623, pairs of \Ox\, emission peaks are symmetrically displaced 
with respect to the central source, in a similar fashion as the high velocity
mm/sub-mm CO/SiO emission associated with molecular jets (e.g. Bachiller et al. 1990; Hirano et al. 2010;
Andr\'e et al. 1990). In IRAS4A, peaks of \Oxline\, emission roughly coincide with peaks of the compact CO(6-5) 
outflow mapped by Yildiz et al. (2012). When compared with H$_2$O, the \Ox\, emission shows remarkable
differences in some of the sources. For instance in VLA1623 and IRAS4A, \Ox\, does not peak
towards the central source, at variance with H$_2$O which peaks on-source in all the objects.
This suggests that, at least for these sources, on-source most of the oxygen has been transformed 
into water by means of high temperature reactions or ice evaporation (Elitzur \& de Jong, 1978). 
Alternatively this could indicate that the jet, traced by OI and not by water, is presently not
particularly active.

The comparison of \Oxline\, with other molecular and atomic species can be seen in detail
in L1448, where maps of several tracers at different wavelengths are available.
Fig. 6 shows an \Ox\, line intensity cut along the outflow 
axis, compared with that of H$_2$O 179$\mu$m, the H$_2$ 17$\mu$m 
and [Fe II]28$\mu$m lines (from Nisini et al. 2013 and Neufeld et al. 2010). 
Spatial resolution of the 17$\mu$m and 28$\mu$m Spitzer data is 11\arcsec\, thus comparable 
to the \Ox\, resolution, limited by the PACS spaxel scale, and to the 
PACS 179$\mu$m diffraction limited resolution of 12$\farcs$7. 
Each line intensity is normalized with respect to its maximum. 
The figure shows that \Ox\, has a spatial profile very similar to that of [Fe II], 
with a main peak on-source and secondary peaks approximately around the B2 and R2 shock clumps. 
H$_2$ does not peak on-source, and the H$_2$O 179$\mu$m emission is slightly less extended 
than \Ox\, with no evident peaks on B2/R2. 
The above evidences suggest that \Ox\, originates from a shock which is at least partially
dissociated, as it has been discussed also in Santangelo et al. (2013)
and Karska et al. (2013).

In L1448 and HH46, where the \Ox\, jet is more extended and the 63$\mu$m line is spectrally resolved,  
we have extracted the \Oxline\, spectra in different positions along the jet,
which are displayed in Fig. 7. In L1448 the radial velocity changes
from blue-shifted to red-shifted, as one is moving from the northern to the
southern position. The same trend is shown in Fig. 8, which presents velocity 
channel maps of the \Oxline\, emission. The velocity increases moving away from the
driving source, a signature which is more prominent in the red-shifted lobe: 
a similar jet acceleration pattern is also observed in molecular 
tracers such as SiO (e.g. Guilloteau et al., 1992) which is a further indication that
 the \Ox\, emission is physically associated with the mm high velocity jet.

In HH46, three spectra have been extracted at the on-source and 
blue/red-shifted jet positions (Fig. 7). The maximum blue- and red-shifted peak velocities
are of the order of -180 and +100 \kms\, (see also van Kempen et al. 2010) which are comparable 
with the velocities
of the atomic jet detected in [FeII] at 1.64\um\, (i.e. from -213 to -177\kms\, in the 
blue lobe and from +92 to +145\kms\, in the red lobe, Garcia Lopez et al. 2010). 
[FeII] emission is also observed at zero velocity, but is not as strong as the \Ox, likely 
due to the extinction. The velocity channel maps are shown in Fig. 9: here the
acceleration of the jet is not so evident as in L1448, since emission from zero
up to high blue-shifted velocity is observed on-source, and the peak of the red-shifted
gas does not shift with velocity. 

\section{Atomic mass flux rates} 

The mass flux rate associated with the \Ox\, jet (\.M$_{jet}$(OI)) is an
important quantity to be estimated in order to assess the relevance of the 
atomic jets in the overall dynamics of the protostellar system. 

Given the high quality of our \Oxline\, maps, and the good, although limited, spatial
and spectral resolution, we can adopt two approaches to estimate the mass loss rates from the
observed emission, the first one based on the \Oxline\, luminosity and the extent of the emission,
and the second one based on shock models. \Oxline\, luminosities have been derived from lines fluxes
reported in Table 2 assuming the distances of Table 1. The computed values are also reported in Table2.
 The comparison between the results from the
two methods will allow us to evaluate the goodness of the assumptions and approximations 
embedded in each approach.

\subsection{\.M$_{jet}$(OI) from line luminosity}

Assuming a steady flow, the mass flux rate can be measured according to the following expression:

\.{M} = $\mu\,m_H\times(n_H\,\mathsf{V})\times V_{t}/l_{t}$,\\

where: $\mu$ is the mean particle weight per H nucleus taken equal to 1.4 (Kauffman et al. 2008), 
 $V_{t}$ and $l_{t}$ are the jet tangential velocity and its length projected
 onto the plane of the sky. This latter is taken as the length of the spatial 
 region considered for the \Ox\, flux extraction, as given in Table 1. 
 The product $(n_H\,\mathsf{V})$ of the hydrogen nuclei
 density and the emitting volume of gas can be related to the  \Oxline\,
 observed luminosity, under the assumption of an optically thin line, as follows:

$n_H\,\mathsf{V} = L(\textrm{[OI]}63\mu \textrm{m})\,(h\,\nu\,A_{i}\,f_{i}\,X[\textrm{O}])^{-1}$.\\

where $A_{i}$ is the spontaneous radiative rate and $f_{i}$ the relative level population.
The quantity $\epsilon$ = $h\,\nu\,A_{i}\,f_{i}\,X[\textrm{O}]$ is the line emissivity
per H nucleus (in erg\,s$^{-1}$), which depends on the kinetic temperature $T_{kin}$ and 
on the density of the colliders $n_{coll}$. $X$(O) is the atomic oxygen abundance, 
which we assume here to be equal to the solar value 4.6$\times$10$^{-4}$ (Asplund et al. 2005).
We remark that the line luminosity depends on which $l_{t}$ scale has been used to
extract the line flux, so that the quantity $ L(\textrm{[OI]}63\mu \textrm{m})/ l_{t}$ does not 
change significantly with the adopted jet length.

To compute the mass flux rate, we therefore need to give constraints 
on the gas physical conditions and on the tangential velocity of the jet.
In principle, we could derive the \Ox\, jet tangential velocity from the observed
radial velocity assuming an outflow inclination angle. However, the limited PACS
spectral resolution and the large uncertainty associated with the velocity
measurement (see section 3) prevent us from estimating $V_{t}$ directly
from the PACS spectra. We instead estimate $V_{t}$ from the proper motion observations
or from radial velocity of the associated high velocity molecular jet.
The adopted values are given in Table 3 and a detailed account on how they have been derived for each source is given 
in Appendix B. The uncertainty on $V_{t}$ is expected to vary between 20 to 100\%, depending on 
how well the jet inclination angle and/or proper motion is known.

\subsubsection{\Oxline\, physical conditions and optical depths effects}

An estimate of the hydrogen particle density is obtained from the \Oxline /[OI]\,145$\,\mu$m 
line ratio, adopting a NLTE code that considers the first five levels of oxygen.
The results are shown in Fig. 10, where this ratio is displayed as a function of 
the density for different values of gas temperatures. 
The observed line ratios obtained from fluxes measured in the jet regions and reported
in Table 2, are here indicated with straight lines.
The considered regions are always larger than the 145$\,\mu$m beam of 11$\arcsec$, hence
we consider negligible flux losses due to diffraction.

We remark that the 63\um/145\um\, ratio does not change significantly, within the errors,
from the on-source position to the positions along the outflows. This indicates 
that we do not appreciate variations of conditions along the jet, and 
also that the 63\um\, line on-source does not significantly suffer from line-of-sight
extinction from the dusty envelope.
The region considered for HH46~IRS has not been mapped in the 145\um\, line: for this source
we therefore consider the 63\um/145\um\, ratio measured in Karska et al. (2013) within a 
single PACS footprint (47$\arcsec \times 47\arcsec$) centered on-source.

Collisions with neutral
and molecular hydrogen are considered: since the fraction of 
H$_2$ molecules surviving in the [O I] emission region is not known, we consider the two
extreme cases of a H gas purely atomic and purely molecular. 
The line ratios observed in the five sources fall in a region of the plot where
the theoretical ratio is rather insensitive to the temperature: however the 
uncertainty on the H$_2$/H ratio translates into an uncertainty of about an order of magnitude
in the derived (H + H$_2$) density, with larger densities obtained for the fully molecular case. 

The emissivity of the \Oxline\, line changes by at most a factor of three 
for kinetic temperatures $T_{k}$ larger than 300 K (see, e.g. Fig. 7 of Liseau 
\& Justtanont 2009). The uncertainty on the colliders density eventually results 
in a factor between 2 and 5 in the line emissivity, for densities less than 
10$^5$ \cmt.

In the above estimate, we have always assumed the \Oxline\, as being optically thin. To
evaluate whether we can disregard opacity effects in the line emissivity
estimates, we have used the Radex radiative transfer code (van der Tak et al. 2007), assuming 
the Large Velocity Gradient (LVG) approximation, 
to calculate for which values of H column density the opacity starts to be
significant. The result is shown in Fig. 11, which plots, as function of density,
the column density of atomic hydrogen at which $\tau$(\Oxline) is equal 1. Here a linewidth of 20\kms\, 
has been assumed. We see that for 
the range of inferred densities, column densities larger than 10$^{22}$ \cmd\,
are required to have an appreciable opacity in the line. Assuming that the \Ox\,
emission originates in a jet with uniform density, the derived $N$(H) limit would
imply, for a density of 10$^4$ \cmt , a jet diameter larger than 0.3 pc, 
which is clearly not consistent with observations. Similar conclusions are reached if one
assumes the colliders to be fully molecular.  

More significant could be the effect of \Oxline\, absorption by the cold gas
along the line of sight, which would not be observed on the spectra due to the 
limited PACS spectral resolution. Such an absorption should not affect
the \Ox\, 145$\,\mu$m, if we assume that the oxygen in the cold foreground gas is mostly populated
in the ground level. 
Absorption by foreground gas was suggested
to explain the very low (i.e. down to $\sim$ 1) \Oxline /[OI]\,145$\,\mu$m line ratios
observed with \textit{ISO}-LWS towards young stars (e.g. Liseau et al. 2006).  
As seen in Fig. 10, our \Oxline /[OI]\,145$\,\mu$m ratios are between 
12 and 20: predictions from shock models give values
between 10 and 35, depending on pre-shock densities and shock velocities 
(Hollenbach \& McKee 1989). Therefore, if foreground absorption is present, it
should affect the observed \Oxline\, intensity by at most a factor of two,
which would increase our density estimates by a factor about $\sim$ 10.
Taking this possibility into account, the combination of increased density
and intrinsic line intensity would result in a decrease of the final mass flux
determination of a factor $\sim$ 2-3. 

In Table 3, the ranges of the mass flux values derived from line luminosity (\.M$_{jet}^{lum}$(OI)) are reported for each jet, 
taking into account the above discussed uncertainties. The uncertainty on the density 
dominates the final error of the \.M$_{jet}^{lum}$(OI). Values range between 10$^{-7}$ and 7$\times$10$^{-6}$ \msol\,yr$^{-1}$, 
with the largest value found in HH46. 

\subsection{\.{M}$_{jet}$(OI) from shock model}

Several lines of evidence suggest that the observed \Ox\, emission has its origin
in dissociative shocks occurring within the jet. 
First of all \Ox\, spatially correlates with [Fe II] (see Fig.6), whose emission is 
associated with dissociative shocks. In fact, C-shocks do not produce a high 
enough degree of ionization to guarantee efficient electronic collisions 
of [FeII] lines (Giannini et al. 2004 ). 
In addition, analysis of \Ox\, lines in several class 0 shock spots
shows that \Ox\, cannot be accounted for by the same non-dissociative 
shock giving rise to the high-J CO and H$_2$O emission (Santangelo et al. 2013, Benedettini et al. 2013).
A similar conclusion on the source position has been found by Karska et al. (2014) from the analysis of the
PACS data of a sample of class 0 and class I sources.

Under the assumption that the \Ox\, emission is connected with dissociative jet shocks, we can estimate the mass loss rate adopting 
models of Hollenbach \& McKee (1989), which predict that \.M
is directly proportional to the observed \Oxline\, luminosity 
provided that the product of the pre-shock density (n$_o$) and shock velocity (V$_{shock}$) is
n$_o \times V_{shock} \la$ 10$^{12}$cm$^{-2}$\,s$^{-1}$ (Hollenbach, 1985):

\.{M}$_{shock}$ = $10^{-4}\,(L([OI]63\mu m)/L_{\odot})$\,\msol\,yr$^{-1}$.

The above relationship assumes that $L([OI]63\mu m)$ is all emitted in 
a single shock caused by the jet/wind launched material impacting on the slow
ambient gas, with V$_{shock} \sim $V$_{jet}$. More realistically, the emission 
we observe is caused by several unresolved shocks along the jet, having 
velocities smaller than the total jet velocity (typical J-shock velocities in
atomic jets from class II sources are of the order of 30-40 \kms\, i.e. about 3-6
times smaller than the jet velocities, e.g. Hartigan et al. 1995). 
Therefore one should in principle correct the above relationship
for these effects, by multiplying it for the term (V$_{jet}$/V$_{shock}\times$(1/$N_{shock}$)), 
where $N_{shock}$ is the number
of shocks in the jet. Under the reasonable assumption that in the considered jet sections 
the number of shocks is of the order of a few, the correction term should be
of the order of unity and thus will not be further considered.

In Table 3, the values of mass flux rates computed from the Hollenbach (1985)
formula (\.M$_{jet}^{shock}$(OI)) are reported and compared with the previous estimates from line
luminosity. These two independent methods give results which are in striking agreement
each other, in spite of the different assumptions.
This may however be not totally fortuitous since both methods are based on the \Oxline\, line 
luminosity. The main difference is that in one case the line theoretical emissivity 
is computed assuming specific physical conditions, while in the shock case
it is assumed that \Ox\, is the main coolant in the post-shock gas 
for temperatures from $\sim$ 5000 to $\sim$ 300 K, and that each particle
entering the shock radiates  in the 63\um\, line about 1.5$\times k\,T$ with $T \sim$ 5000 K.

\section{Discussion}

The PACS observations reveal the presence of atomic collimated emission 
associated with the outflows of five class 0 sources. We wish to examine the
role of this emission in the global energy budget of the protostellar system. 
In particular, we can estimate the \.M$_{jet}$(OI)/\.M$_{acc}$ ratio, with \.M$_{acc}$
the mass accretion rate onto the protostar,  to 
evaluate the efficiency of the atomic jet in removing matter from the accreting
system. At the same time, we can also discuss if this atomic jet possesses enough momentum to sustain the 
entrained large scale and powerful CO outflows associated with the considered sources.  

Measurements of \.M$_{acc}$ in class 0 sources are 
difficult and have a large uncertainty,
because the accretion region is obscured from direct view by the optically thick
massive envelope and because the already accumulated stellar mass is generally unknown. 
A rough estimate of \.M$_{acc}$ can be obtained by assuming
that the source bolometric luminosity is dominated by accretion and that the mass 
of the central star is 0.2 \msol . This latter assumption is supported by recent 
interferometric observations of rotating disks in VLA1623 and similar class 0 sources,
inferring stellar masses around 0.2 \msol\, (Murillo et al. 2013,
Codella et al. 2014). Taking the bolometric luminosities given in Table 1
and further assuming $R_{\star} \sim$ 4\,R$_\odot$, we derive 
mass accretion rates ranging from 0.5 to 12$\times$10$^{-6}$  \msol\,yr$^{-1}$. 
Figure 12a shows the derived \.M$_{acc}$ as a function of
the \.M$_{jet}$(OI), where both determinations, from  \Oxline\, luminosity
and the shock model, have been considered (blue triangles and open red circles, respectively).
\.M$_{jet}$(OI)/\.M$_{acc}$ ratios is in the range 0.05-0.5. 
These values are in line with current MHD jet launching models (e.g. Ferreira et 
al.2006), where mass loss rates are
predicted to be $\sim$ 0.05-0.5 of the mass accretion rate
and they are also consistent with the \.M$_{jet}$(OI)/\.M$_{acc}$ ratios measured in more
evolved class I and CTT sources for which values in the range 0.05-0.4 are typically
found (Ellerbroek et al. 2013, Antoniucci et al. 2008). 

The role of the \Ox\, jet on the overall outflow dynamics can also be addressed by comparing
the \Ox\, mass flux with estimates of the outflow thrust (or force, $F_{CO}$). This
latter is usually inferred from millimeter observations of the CO emission. Derivation of $F_{CO}$ is subject to
a large uncertainty due to the different methods used to determine
outflow mass and the different corrections for observational
effects, such as opacity and inclination (see e.g. Downes \& Cabrit, 2007, 
van der Marel et al. 2013).
 If the outflow is driven by the underlying
jet with momentum conservation, then the (total) jet mass loss rate and outflow
thrust should be related by  \.{M}$_{jet}$(CO) = $F_{CO}$/V$_{jet}$. V$_{jet}$ is the total
jet velocity that can be estimated from the jet radial velocity and inclination
angle following the same considerations adopted for the tangential velocity in Appendix B.  
Literature determinations of \.{M}$_{jet}$(CO) are reported in Table 3 for each outflow; 
the corresponding range of values in some
cases spans up to an order of magnitude.

The resulting \.M$_{jet}$(OI) vs \.M$_{jet}$(CO) plot is shown in Figure 12b:
we note here that for two of the sources, namely NGC1333-IRAS4A and HH46 IRS, 
the two mass loss determinations are consistent within the errors, indicating that the \Ox\, jet 
may have a role in driving the large scale outflow. For the other three sources, however, 
\.M$_{jet}$(OI) is more than an order of magnitude smaller than the value needed 
to sustain the outflow, even taking into account the large  \.M$_{jet}$(CO) uncertainty.
Comparing the results in Figs. 12 a and b, it appears clear that the mass flux derived 
from CO in L1448, VLA1623 and BHR71
is in contradiction with the \.M$_{acc}$ estimated from their bolometric luminosity,
since it implies \.M$_{jet}$(CO)/\.M$_{acc}$ $\ge$ 1. 

Therefore for these three sources either the \.M$_{jet}$(CO) is largely overestimated or
\.M$_{acc}$ is underestimated, due for example to a central mass lower  
than assumed. In this latter case, \.M$_{jet}$(OI) in these three sources would be $<$ 0.01 \.M$_{acc}$, 
thus insufficient both to account for
driving the outflow and to have a role in limiting the accreting material onto the star. 

Of the three sources, BHR71 has more uncertainty in the \.M$_{jet}$(CO) estimation, 
because the inclination of the outflow is poorly known and because there is the contamination 
by the secondary outflow from IRS2. 
In the case of the VLA1623 outflow, the larger error in the thrust determination
comes from disentangling the outflow gas at low radial velocity from the envelope gas
because of the low outflow inclination with respect to the plane of the sky. 
Andr\'e et al. (1990), however,  estimate that the error 
due to this effect would not change the outflow mass by a factor more than two.  

Finally, the outflow force in L1448 has been measured
by different authors adopting different data and methods. In particular, the 
value originally measured with single dish, low-J lines
observations (Bachiller et al. 1990), has been confirmed with the analysis of high-resolution
interferometric (Hirano et al. 2010) and high-J CO (Gomez-Ruiz et al. 2013, Y{\i}ld{\i}z et 
al.(2014)) observations.
It is therefore unlikely that all those determinations are overestimated by a factor 
of 10. In order to reconcile the high CO mass ejection rate in L1448 with its
low bolometric luminosity, Hirano et al. (2010) assume that the protostellar mass should be in the range 0.03 - 0.09 \msol . 
Time variability in the disk accretion rate could also be at the 
origin of the found discrepancy. 

We note that the CO outflow of L1448 is peculiar with respect to the others, since it presents
Extreme High Velocity (EHV) molecular gas in the form of CO/SiO collimated emission,
corresponding to de-projected velocities of the order of about 200\kms\, (Bachiller et al. 1990, Hirano et al. 2010).
Interferometric mm observations of this emission suggest that 
it may be directly linked to the original protostellar jet and not to entrained
material. This is also supported by the very different chemical composition of the 
EHV gas compared to the lower velocity outflow gas, pointing to a different origin
of the two components (Tafalla et al. 2010). 
The above findings suggest that the primary jet responsible for driving the large scale low velocity outflow
of L1448 is mainly molecular and that the atomic gas contributes only
for a smaller part to the total jet momentum. Dionatos et al. (2007) discovered for the first 
time the atomic jet in L1448 with \textit{Spitzer}, through its [S I] and [Fe II] mid-IR emission. 
The mass flux derived for this atomic emission,
detected through single pointing IRS observations, was in the range 
2-20$\times$10$^{-7}$ \msol\,yr$^{-1}$, which is consistent with the values derived here from [OI]. 
An evaluation of the relative contribution of the atomic and molecular jet to the dynamics of the
outflow can be done in L1448, since Hirano et al. (2010) computed the CO force separately for the EHV jet and for the outflow
low velocity shell. The $F_{CO}$ of the molecular jet is $\sim$\,5$\times$10$^{-4}$ \msol\,km\,s$^{-1}$\,yr$^{-1}$,
implying a mass flux rate of $\sim$\,4$\times$10$^{-6}$ \msol\,yr$^{-1}$ for a jet velocity of 180\kms, which is
an order of magnitude higher than \.M$_{jet}$(OI). $F_{CO}$ of the slow swept out gas is instead of 
the order of 1.8$\times$10$^{-3}$ \msol\,km\,s$^{-1}$\,yr$^{-1}$, i.e. about a factor of three higher than 
the force of the jet. Given the involved uncertainties, the EHV jet could indeed be the main
agent in driving the slow speed shells.

No CO jet-like emission has been mapped in VLA1623 and BHR71, due to the lack of high resolution 
observations: it is however remarkable that they both show evidence of EHV gas with deprojected velocities 
in excess of 70\kms\
(Kristensen et al. 2013, Andr\'e et al. 1990). This may suggest that also in these two sources the discrepancy between 
\.M$_{jet}$(CO) and \.M$_{jet}$(OI) could be due to the fact that the atomic \Ox\, jet represents only a
minor component of the original jet and that this latter is mainly composed of molecular
material. 

HH46, is the more evolved of the sampled sources, being classified between the class 0 and class I stage 
(Antoniucci et al. 2008), and has an already well
developed atomic jet observed in high excitation optical and near-IR forbidden lines.
The mass flux rate of the jet has been estimated to be in the range 0.5-1$\times$10$^{-7}$ \msol\,yr$^{-1}$
from the near-IR [Fe II] emission (Garcia Lopez et al. 2010). This value is smaller by a factor of at least 20 
compared to that inferred from \Oxline.  This reinforces the result of Fig. 12, 
i.e. that in this source the large scale outflow is mainly sustained from the low excitation neutral atomic jet traced by the \Oxline\, emission.

The above findings are in line with the idea that jets undergo an evolution in their composition. 
Jets in the youngest sources remain mainly molecular while the atomic component progressively 
becomes dominant, as the jet gas excitation and ionization increases.
Jets from more evolved pre-main sequence T Tauri
stars are in fact bright in UV-optical lines showing larger ionization fractions, 
($x_e \sim$ 0.2-0.5) and electron temperatures ($T_e \sim$ (0.5-2)\,10$^4$ K) with respect to younger class I sources 
(having $x_e \sim$ 0.05-0.1 and $T_e \sim$ 10$^4$ K, Nisini, 2009). 
Although PACS observations of a few T Tauri stars indicate that the \.M$_{jet}$ derived from \Oxline\, is 
higher or comparable to the value derived from optical lines (Podio et al 2012), these values
should be considered as upper limits as \Ox\, emission in these evolved sources could be contaminated
by PDR emission (e.g. Karska et al. 2014).
 Such an evolution of physical parameters and composition is for example predicted in disk winds 
models where irradiation from X-ray and far-UV photons from accretion hot spots are taken into
account (Panoglou et al. 2012). 
 
Although the statistics are small, and should be confirmed on larger samples of sources, 
our results suggest that the contribution of warm atomic gas on the overall
jet composition is dominant only in an intermediate phase of jet evolution, 
after the originally molecular gas is getting progressively dissociated (early class 0), and before 
the jet becomes mainly hot and ionized (CTTs).

\section{Conclusions}

We have carried out mapping observations of the \Oxline\, and 145\um\, lines towards protostellar outflows in the  
L1448, NGC1333-IRAS4, BHR71, VLA 1623 and HH46 star forming regions using the PACS instrument on 
{\it Herschel}.  
We have used these maps to detect atomic jets at low excitation, that could be the responsible 
agent for driving the large scale molecular outflows, and to study their dynamical properties.

The main results obtained from this work are the following:

\begin{itemize}
\item We observe, close to the driving sources, spatially resolved [O I] emission extended along the outflow direction,
that we attribute to the presence of a collimated jet. This interpretation is reinforced in three
out of the five considered sources (namely, L1448, BHR71 and HH46), by the kinematics of the \Ox\, emission: we were in fact able to separate the red- and blue-shifted jet lobes, having maximum velocities 
up to 200\kms. 
\item In L1448 and HH46, having the largest velocity spreads, velocity channel maps 
have been constructed in order to study their variations as a function of distance from
the source. The clearest pattern is shown by the \Ox\, emission in L1448, where we observe that 
the velocity increases with distance from the driving source: a similar acceleration has been 
previously observed in the SiO collimated jet associated with this source, suggesting that the \Ox\, and molecular
gas are kinematically connected and that this latter could represent the colder cocoon of 
a jet at higher excitation. 
\item Complementary observations of the \Ox\,145\um\, line are used to constrain the gas density
(n(H+H$_2$)) of the atomic jet between 10$^3$ and 10$^5$ \cmt\, assuming T$_k \ga$ 300 K. 
We additionally infer that self-absorption of the \Oxline\, should not be significant, while absorption by
cold gas along the line of sight should affect the observed line flux by at most a factor of two.
\item Mass flux rates for the atomic jet (\.M$_{jet}$(OI)) have been measured from the \Oxline\, luminosity,
adopting two independent methods. The first one is based on the \Oxline\, luminosity and emission extent,
and needs to assume the gas physical conditions and tangential velocity, while the second one is based 
on the assumption that the emission comes from the dissociative shock occurring when the primary jet/wind
interacts with the ambient medium. The two determinations are consistent with each other and
provide values in the range (1-4)\,10$^{-7}$ \msol\,yr$^{-1}$ for L1448, IRAS4A, BHR71 and
VLA123, while an order of magnitude higher value for HH46. 
\item The derived mass flux rates are compared with mass accretion rates (\.M$_{acc}$) to the central protostar and the 
\.M$_{jet}$(CO) estimated from the large scale cold CO maps, assuming momentum conservation in the
interaction between the jet and the ambient medium. 
We find \.M$_{jet}$(OI)/\.M$_{acc}$ ratios 
in the range 0.05-0.5, similar to values found in more evolved class I/II sources. 

The \.M$_{jet}$(OI) is comparable to \.M$_{jet}$(CO) in only two sources, namely HH46 and IRAS 4A, 
while it is more than an order of magnitude smaller than \.M$_{jet}$(CO) in the remaining three, L1448, VLA1623, and BHR71. 
Since these three sources with smaller \.M$_{jet}$(OI) are associated with the molecular gas at extreme velocities, 
we speculate that most of the mass flux is carried out by molecular jet and that the contribution of the atomic 
jet to the acceleration of the large scale outflow is less significant.

\end{itemize}

The above findings suggest an evolution in the properties
of jets during the protostellar and pre-main sequence lifetime. 
In this scenario, the contribution of warm atomic gas, traced by the \Oxline\, line, to the overall
jet composition is dominant only in an intermediate phase of jet evolution, 
after the originally molecular jet gets progressively dissociated (early class 0), and before it
becomes mainly hot and ionized (class II). \Oxline\, mapping of jets from more class 0 and class I sources
 would be needed to increase statistics on the role of the warm atomic gas
in the dynamics of protostellar jets. In this respect, observations with the SOFIA-GREAT instrument will provide the 
spectral resolution high enough to obtain velocity-resolved [OI] maps in a larger sample
of sources, allowing one to study  the dynamics of the [OI] protostellar jets in much more details. 

\appendix

\section{Outflow descriptions and \Ox\, general morphology}

\textbf{L1448}: 
located in the Perseus molecular cloud (\textit{D} = 232 pc, Hirota et al. 2011), the 
low luminosity class 0 source L1448-C (or L1448-mm) drives a powerful 
and extended molecular outflow (Bachiller et al. 1990) that interacts in the northen region, 
with two other compact flows originating from the small three star cluster L1448-N (Looney et al. 2000).
L1448-C itself is in fact a binary (CN and CS, J{\o}rgensen et al.(2006)) with the CN source 
responsible for the excitation of this extended outflow.
Close to the driving source, interferometric CO and SiO observations have resolved the
outflow into two kinematically distinct components; a very collimated and high velocity 
component seen in both CO and SiO, and low velocity cavity walls only seen in CO
(Hirano et al. 2010). Warm molecular gas, with temperature in excess of $\sim$ 300 K, has been probed by
different near- and far-IR observations of H$_2$, CO and H$_2$O all along the outflow
(Davis \& Smith, 1996; Neufeld et al. 2009; Giannini et al. 2011; Nisini et al. 2000; Nisini et al. 2013; Santangelo et al. 2013).
\textit{Spitzer} observations have provided evidence for the presence of an atomic jet at low excitation, 
emitting in [SI], [FeII] and [SiII] lines (Dionatos et al. 2009).
The \Oxline\, map (Fig.1) shows bright emission peaks towards the sources
of the field. In particular, elongated jet-like emission is detected towards L1448-CN/S,
while the strongest peak is located at the L1448-NA source. Additional 
peaks are observed along the outflow in correspondence with known shock spots B2/R2 R3 and R4. 

\textbf{NGC1333-IRAS4}: 
NGC1333-IRAS4 consists of four very embedded objects located in the L1450 dark cloud within the Perseus
molecular cloud (D = 235 pc, Hirota et al. 2008). IRAS4A and B are separated by about 30\arcsec, and each of
them is a binary (A and A', separated by 2\arcsec, and B and B', separated by 11\arcsec\, Lay et al. 1995, Looney et al. 2000).
Outflows have been detected in both the A and B sources, but with a very different morphology. The 4A 
outflow extends more than 4\arcmin, has been detected in both CO and SiO, and shows a remarkable 
directional variability, with the position angle quickly changing from 0 
to about 45\arcdeg\, close to the source, and again showing a sharp bend in the middle of the northeastern lobe
(Choi et al. 2006). At variance, the IRAS4B flow is very compact, extending only for about 20\arcsec\,
from source (J{\o}rgensen et al. 2007, Plunkett et al. 2013). Herschel detected strong 
warm molecular emission from CO and H$_2$O associated with the outflows of both IRAS4A and B (Herczeg et al.
2012; Santangelo et al. 2014; Karska et al. 2013).
 
The PACS map presented in Fig. 3 covers the outflows for both IRAS4A and 4B. The
outflow from IRAS4B is very compact and the \Oxline\, emission is offset with respect
to the central source, as discussed in Herczeg et al. (2012). Emission is detected along
the extended IRAS4A outflow with a strong peak on the shocked R2 position, where also
H$_2$O and high-J CO have been detected by \textit{Herschel} (Santangelo et al. 2014).

\textbf{HH46}: 
HH46 IRS is located in an isolated Bok globule close to the Gum nebula (at a distance of 450 pc).
It is an highly accreting object (Antoniucci et al. 2008) driving a well known and spectacular outflow 
firstly discovered at visible and IR wavelengths
(Heathcote et al. 1996, Hartigan et al. 2005, Eisl\"offel \& Mundt 1994). 
Optically, the brightest part of the flow is the blue-shifted northeastern region, where
strong HH objects have been detected with the Hubble Space Telescope. The HH46 IRS red-shifted 
counter-jet, moving at radial velocities up to 150\kms , has been detected in the infrared 
from both ground-based and \textit{Spitzer} observations (Garcia Lopez et al. 2010, Velusamy et al. 2007).
HH46 IRS also drives an extended molecular outflow (Chernin \& Masson 1991, van Kempen et al. 2009),
that has been recently studied at high resolution with ALMA CO(1-0) observations 
(Arce et al. 2013). The morphology and kinematics of the molecular gas supports the presence
of a wide angle wind producing an extended cavity, plus a collimated episodic flow. 
Far-IR line emission from H$_2$O, CO, \Ox\, and OH have been observed both on-source
and on the outflow by Herschel (van Kempen et al. 2010).

The \Oxline\, map presented here (Fig.5) covers the full extent of the outflow up to the two bright
terminal bow shocks  HH47A and HH47C (e.g. Eisl\"offel \& Mundt 1994). 
The emission is centrally peaked and extends along the outflow with different emission knots.
The two emission peaks at the extreme of the map are located just ustream of HH47A and HH47C. 
This spatial offset between the \Oxline\, emission and the HH objects location is probably 
just due to a poor sampling at the map edges. 

\textbf{BHR71}: 
BHR71 is a close-by isolated Bok globule (D$\sim\,$200 pc), hosting an extended bipolar outflow
originating from a binary protostellar system (IRS1 and IRS2, Bourke et al. 1997, Bourke 2001). 
Further investigations have shown that both sources are driving an outflow: IRS1 is responsible
for the extended flow in the north-south direction, while IRS2, the lower luminosity source,
drives a more compact flow having a P.A. of $\sim$-30 degree (Parise et al. 2006, Yildiz et al. 2014).
The direction of the two outflows is indicated with arrows in Fig. 2.
Weak optical HH objects (HH320 and HH321) have also been detected along the flows (Corporon \& Reipurth 1997). 
The region mapped by our PACS observations covers the central 4\arcmin\, part of the outflow,
that has a length of more than 8\arcmin\, ($\sim$ 0.4pc). The same region investigated here has
been mapped with \textit{Spitzer-IRS}, showing bright emission from H$_2$ mid-IR rotational 
lines (Neufeld et al. 2009, Giannini et al. 2011).
From Fig. 2 we see that the \Oxline\, has a strong peak towards the BHR71-IRS1 source while a much
fainter emission is associated with the IRS2 source. Emission peaks are also detected
on the cavity walls of the IRS1 and IRS2 outflows, also associated with the HH320 and 321 objects.

\textbf{VLA1623}: 
The collimated outflow firstly discovered by Andr\'e et al. (1990) in the $\rho$ Oph core A (D = 120 pc, Lombardi et al. 2008),
is driven by the first identified class 0 source object VLA 16234-2417. The source itself was
recently discovered to be composed of three different protostellar condensations, two of them, VLA1623 A and B,
separated by only 1\farcs 2 (Looney et al. 2000, Maury et al., 2012) and a third faint millimeter
source (VLA1623W) at about 10\arcsec\, (Murillo et al., 2013). 
Along the VLA1623 outflow, several near-IR H$_2$ and knots have been detected (Dent et al. 1995), as well
as optical HH objects (HH313), located close to the reflection nebula of the YSOs  GSS30 (Wilking et al. 1997).
A bright nebula, illuminated by the B3 star S1, is located north of VLA1623, and dominates 
the mid-IR emission of the region with PAH features (Liseau \& Justtanont, 2009). A bright photo-dissociation
region (PDR) is also associated with this nebula, and emits strongly in \Ox\, and [C II] far-IR fine
structure lines (Liseau et al. 1999).

The \Oxline\, map (Fig. 4) covers the VLA1623 outflow and part of the PDR, this latter dominating the \Ox\, emission
in the  north-east direction. Collimated \Oxline\, emission is observed close to the source.
Emission is also detected all along the outflow, but the strongest
peaks are observed in the weastern lobe, where the HH313A and B objects and the GSS30 source are located.

\section{Tangential velocity}

In this appendix we briefly summarize the derivation of the
jet tangential velocity (V$_t$) adopted for the mass flux determination for each source . These
values are reported in Table 3.
 \textbf{L1448:} the maximum
radial velocity of the SiO high velocity jet is 65\kms\, (Guilloteau et 
al., 1992) and the jet inclination angle has been estimated as 21\degr\, (Girart \& Acord, 2001) with 
respect to the plane of the sky. This
implies a tangential velocity of 170\kms. Given the similarity of the \Ox\, 
morphology and kinematics with the molecular jet, we assume the same velocity for
the \Ox\, jet. 

\textbf{IRAS4:} Choi et al. (2006) measured an outflow proper motion
of about 100\kms\, from H$_2$ multi-epoch emission. This could be a lower
limit to the \Ox\, jet velocity, since proper motion has been measured on shock
knots far from the central source, and thus not directly associated with the jet. 
However, CO observations on source show that the maximum CO radial velocity is
of the order of 25\kms: even assuming an inclination angle of only 10\degr ,
this implies a maximum tangential velocity of 140\kms. \textbf{HH46:} Eisl\"offel \& Mundt (1994) measure a proper motion of the HH46 atomic jet of 200$\pm$100 \kms\, and a jet
inclination angle of about 30\degr . Garcia Lopez et al. (2009) measure a [FeII] 1.64\um\,
radial velocity in the inner jet of about 200\kms . This implies a $V_{t}$ of 300\kms, 
assuming $\theta$ = 30\degr . \textbf{BHR71:} the inclination angle of this outflow
is quite uncertain and has been estimated in the range 25-65\degr . Our \Ox\, PACS
map shows that the blue- and red-shifted lobes of the atomic jet are quite well
separated, favouring an inclination angle of no more than 45\degr . The maximum
radial velocity of the CO high velocity gas is about 50\kms, therefore we estimate 
a tangential velocity in the range 50-110\kms. \textbf{VLA1623:} Davis et al. (1999)
estimate an inclination angle of 15\degr\, for this outflow . 
The CO outflow (Andr\'e et al. 1990) presents wings up to a maximum of 15\kms\,
in the red-lobe, that, assuming the inclination of 15\degr , corresponds to a maximum 
tangential velocity of 56\kms\ . This is consistent with the proper motion measurements
by Caratti o Garatti (2007) from H$_2$ observations, that found a tangential velocity in the red-shifted jet of 74$\pm$15\kms.


\begin{table}
\caption{Observed outflows}             
\label{table:1}               
\begin{tabular}{l l l l l}     
\hline\hline       
Source & R.A. (J2000)$^{a}$ & DEC (J2000)$^{a}$ & D & $L_{bol}^{b}$\\
&  $h$ $m$ $s$ & \degr ~\arcmin ~\arcsec & pc & L$_\odot$\\
\hline                    
L~1448 & 03 25 38.79 &  +30 44 09.32 & 232 & 9\\
NGC~1333-IRAS4 &03 29 10.58 & +31 13 24.37 & 235& 9.1\\
HH~46 & 08 25 43.90 & -51 00 36.0 & 450 & 27.9\\
BHR~71 & 12 01 35.44 & -65 08 55.97 & 200 & 14.8\\
VLA~1623 & 16 26 26.46 & -24 24 24.78 & 120 & 1.1\\

\hline                  
\end{tabular}
\\
$^{a}$ Coordinates of the central map position\\
$^{b}$ From Kristensen et al. 2012 except VLA1623 where $L_{bol}^{b}$ is taken from Murillo et al. (2013)\\
\end{table}

\begin{table}
\caption{Parameters of \Ox\ spectra extracted in the jet regions.}           
\label{table:1}      
\begin{tabular}{ccccccc}     
\hline\hline   
       &        & \multicolumn{4}{c}{63\um} & 145\um \\ 
     \cline{3-6} 
Source & Region & $F\pm \Delta F$ & $V_{LSR}$ & $\Delta V$ & $L/L_{\odot}$ & $F\pm \Delta F$\\

\hline
L~1448 & 16\arcsec $\times$ 45\arcsec, P.A. = 165$^o$ & 11.0$\pm$0.1 & $+$18 & 170 & 1.8$\times$10$^{-3}$ & 0.6$\pm$0.1 \\
NGC~1333-IRAS4A & 20\arcsec $\times$ 38\arcsec, P.A. = 0  & 5.3$\pm$0.6 & $+$56 & 130 & 9.1$\times$10$^{-4}$& 0.5$\pm$0.1\\
HH~46 & 32\arcsec $\times$ 59\arcsec, P.A. = 45 & 32.0$\pm$0.6 & $+$32 & 184 & 2.0$\times$10$^{-2}$& 2.3$\pm$0.1\\
BHR~71 & 33\arcsec $\times$ 33\arcsec, P.A. = 0 & 26$\pm$0.1 & $-$53 & 160 & 3.2$\times$10$^{-3}$ & 2.0$\pm$0.4\\
VLA~1623 & 19\arcsec $\times$ 78\arcsec, P.A. = 125$^o$ & 46$\pm$0.5 & $+$44 & 125 & 2.1$\times$10$^{-3}$ & 2.2$\pm$0.5\\
\hline
\end{tabular}
\\
Note - The considered rectangular regions are centered on-source.
Flux $F$ in units of 10$^{-13}$ erg\,s$^{-1}$\,cm$^{-2}$. [OI]63$\mu$m luminosity computed from reported fluxes assuming
distances in Table 1. $V_{LSR}$ and $\Delta V$ 
in \kms . $\Delta F$ is the 1$\sigma$ statistical error: an additional flux calibration
error of $\sim$ 20\% needs to be considered. $\Delta V$ is the observed line FWHM, not deconvolved for the instrumental
profile.

\end{table}

\begin{table}
\caption{\Ox\, mass flux rate estimates and comparison with \.M$_{jet}$(CO) and \.M$_{acc}$}             
\label{table:1}      
\begin{tabular}{cccccccc}     
\hline\hline       
Source & $n$(H+H$_2$)& $V_t$ & \.{M}$_{jet}^{lum}$(OI) $^a$ & \.{M}$_{jet}^{shock}$(OI) $^a$& \.{M}$_{jet}$(CO)$^b$& Ref. $^c$ & \.{M}$_{acc}$ $^d$ \\
       & 10$^4$\cmt & \kms & \multicolumn{3}{c}{10$^{-7}$ \msol\,yr$^{-1}$}  & & 10$^{-6}$ \msol\,yr$^{-1}$\\
\hline
L~1448 & 1-10& 170 & 2-4 & 2-4 & 27-110 & 1/2/3/4 & 3.7\\
IRAS4A & 0.1-3 & 100-140 & 1.2-11& 1-2 & 1.5-38 & 4/5/6 & 3.8\\
HH~46 & 0.4-6 & 300 & 22-88& 20-40 & 15-28 & 4/7 & 12\\
BHR~71 & 0.2-3 & 50-100 & 2-4 & 3-6 & 42-530 & 4/8 & 6.2\\
VLA~1623 & 2-20 & 60 & 1-4 & 2-4 & 16-160 & 9 & 0.5\\
\hline                  
\end{tabular}
\\
$^{a}$ \.{M}$_{jet}^{lim}$(OI) is the mass flux rate measured from the \Oxline\, luminosity following
the procedure in Sect. 3.2. \.{M}$_{jet}^{shock}$(OI) is the mass flux rate derived 
following Hollenbach (1985).\\
$^{b}$ \.{M}$_{jet}$(CO) is the mass loss rate in the original jet/wind needed to drive the CO large scale outflow assuming
momentum conservation and it has been estimated as \.{M}$_{jet}$(CO) $\times$ V$_{jet}$ = $F_{CO}$, where $F_{CO}$
is the measured outflow force in \msol\,yr$^{-1}$\,\kms  \\
$^{c}$ References for the outflow parameters are: 1) Bachiller et al. (1990); 2) Hirano et al. (2010); 
3) Gomez-Ruiz et al.(2013); 4) Y{\i}ld{\i}z et al. (2014); 5) Knee \& Sandell (2000); 6) Blake et al. (1995); 7) Van Kempen et al. (2009); 
8) Bourke et al. (1997); 9) Andre' et al. (1990)\\
$^{d}$ \.{M}$_{acc}$ estimated assuming that L$_{bol}$ given in Table 1 is dominated by the accretion 
luminosity, and that M$_*$ = 0.2 \msol \\

\end{table}

\clearpage



\begin{figure}
\epsscale{1.2}
\plottwo{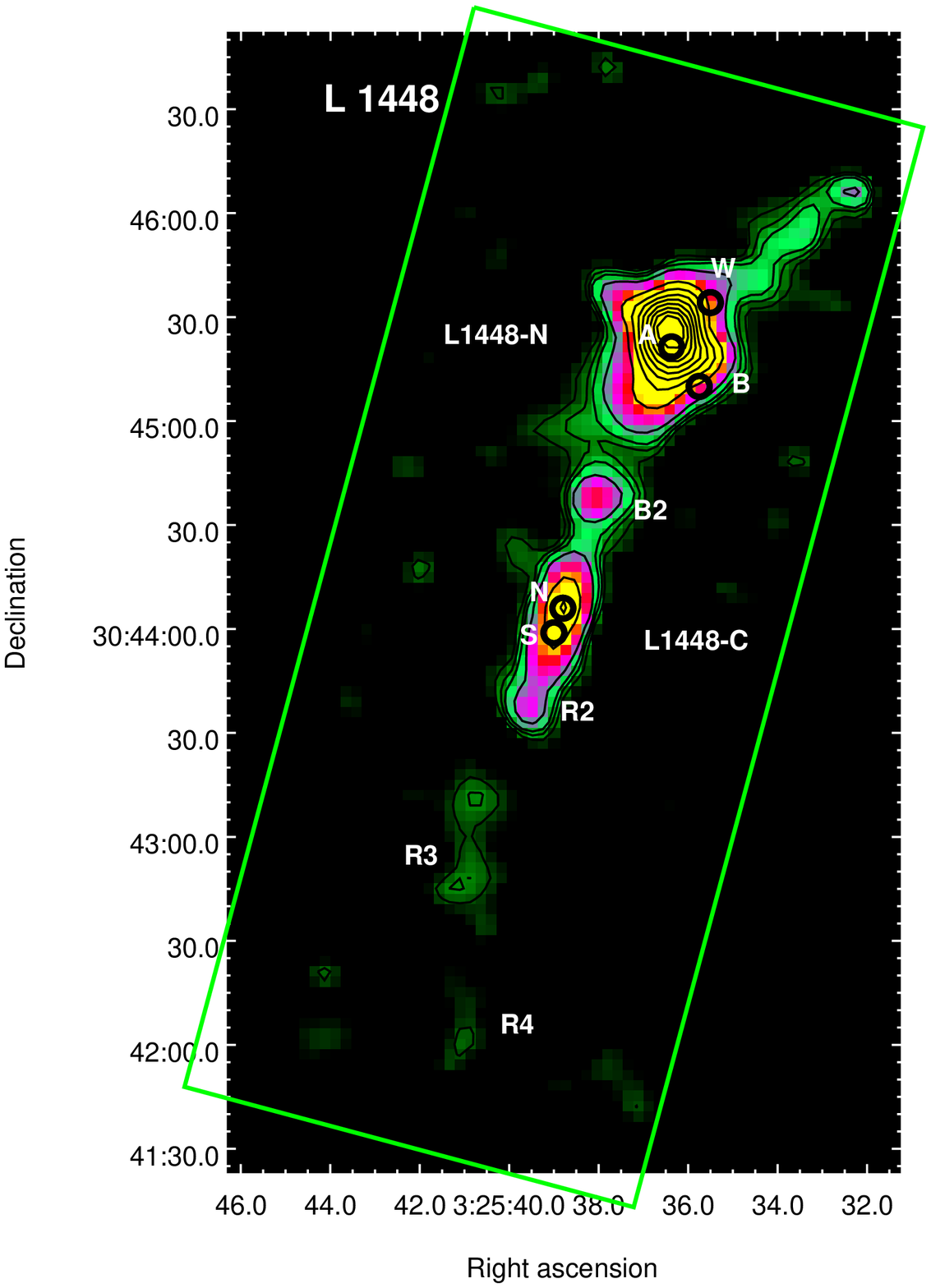}{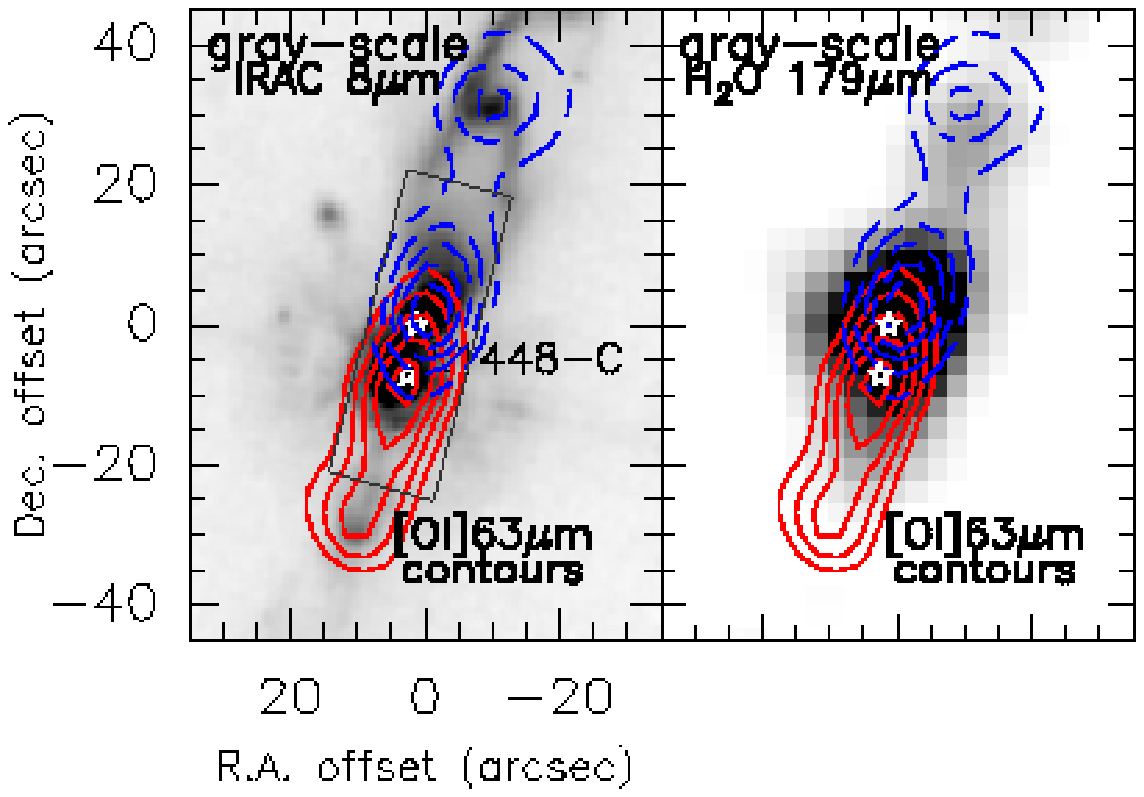}

\caption{\textbf{Left:} continuum subtracted PACS map of the \Oxline\, towards L1448. Labels indicate
the positions of the protostars in the region, namely L1448 C-N and C-S, at the center, and L1448 N-A, N-B and N-W, 
in the northern region, as well as the main emission peaks following the nomenclature for CO of Bachiller et al. (1990).
The green rectangle delimits the region mapped with PACS.
\textbf{Right:} enlargement of the L1448-C central region, where 
blue- and red-shifted \Oxline\, emission is displayed with  dashed and solid contours, respectively. 
V$_{LSR}$ integration limits are -200 to 0 \kms\, and 0 to +200 \kms\, for the blue- and red-shifted emission, respectively. Contours are drawn from 2\,10$^{-5}$ to 6\,10$^{-5}$ \escs\, in steps of 1\,10$^{-5}$ \escs. The contours are overlayed on
a gray-scale \textit{Spitzer}-IRAC archival image (left), and on a  PACS map of the \wat\, 179\um\, line 
(from Nisini et al., 2013). The gray box delineates the region where the jet integrated flux has been 
extracted. \label{1448}}
\end{figure}

\begin{figure}
\epsscale{0.8}
\plotone{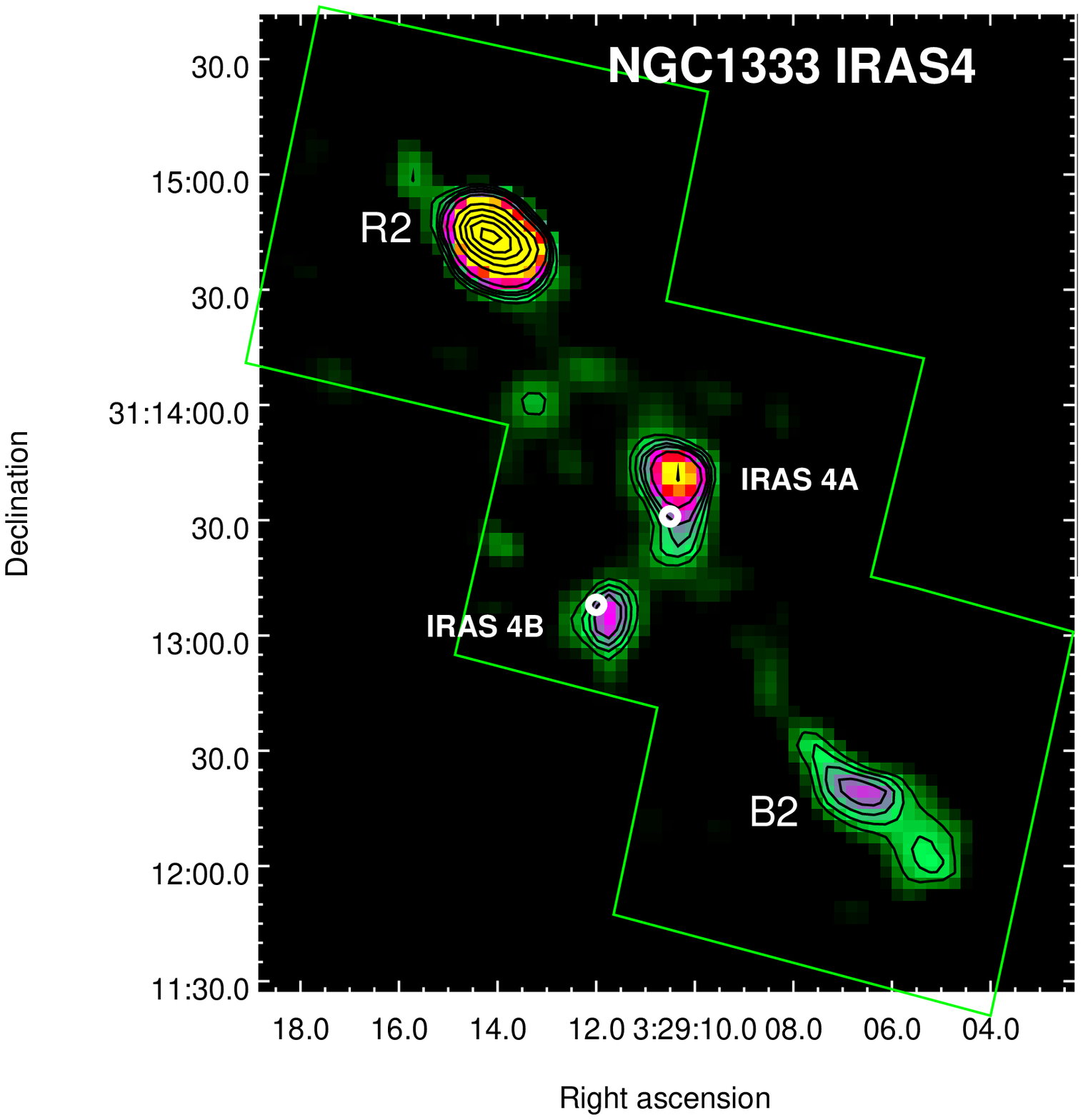}
\epsscale{0.7}
\plotone{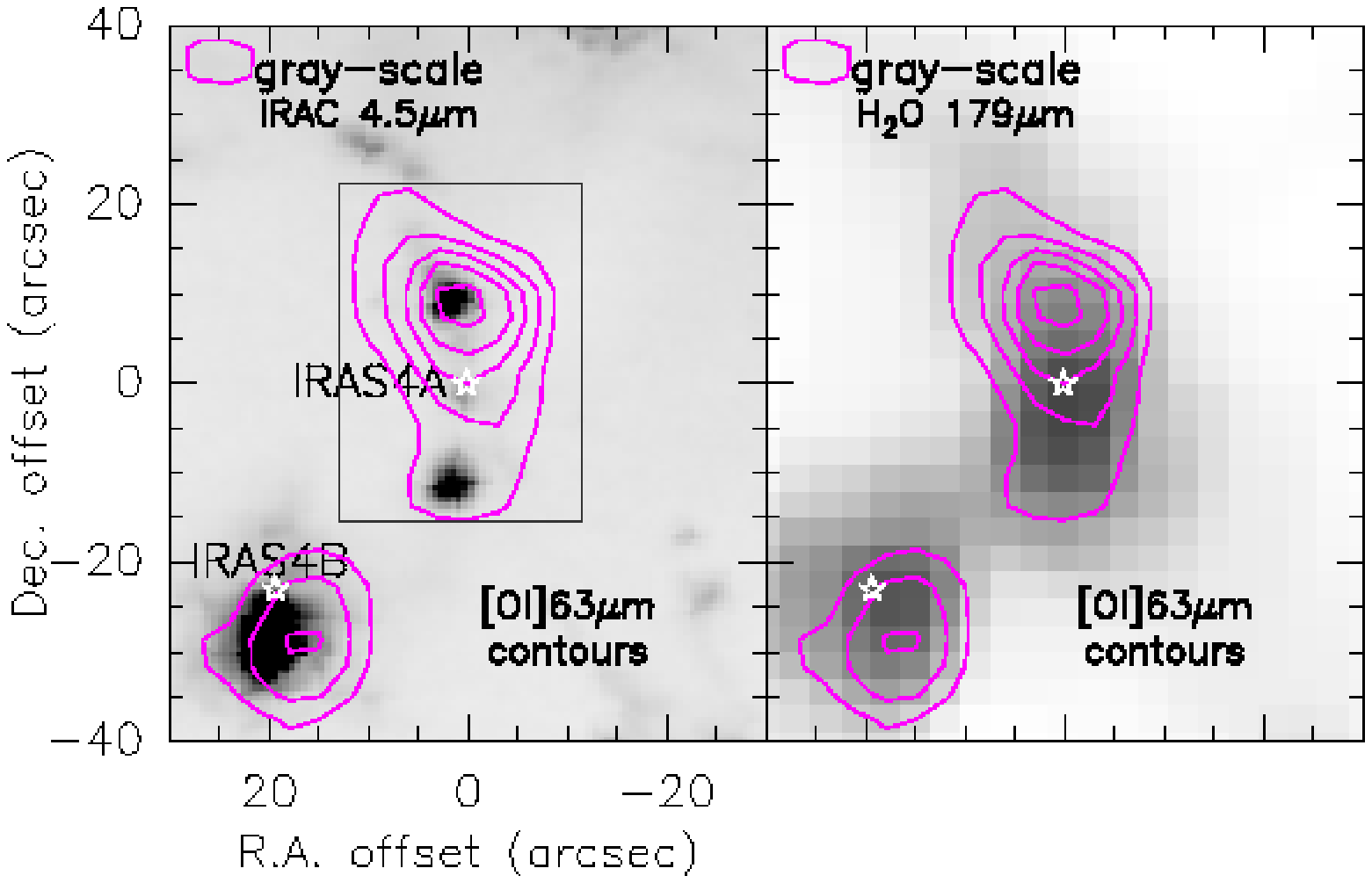}
\caption{As in Fig. 1 for IRAS4A and B (but with the 4.5\um\, Spitzer image in place 
of the 8\um\, image). Contours refer to
the total integrated line intensity and run from 1.5\,10$^{-5}$ to 7.5\,10$^{-5}$ \escs\,
in steps of 1.5\,10$^{-5}$ \escs\,\label{fig1}}
\end{figure}

\begin{figure}
\epsscale{1.}
\plotone{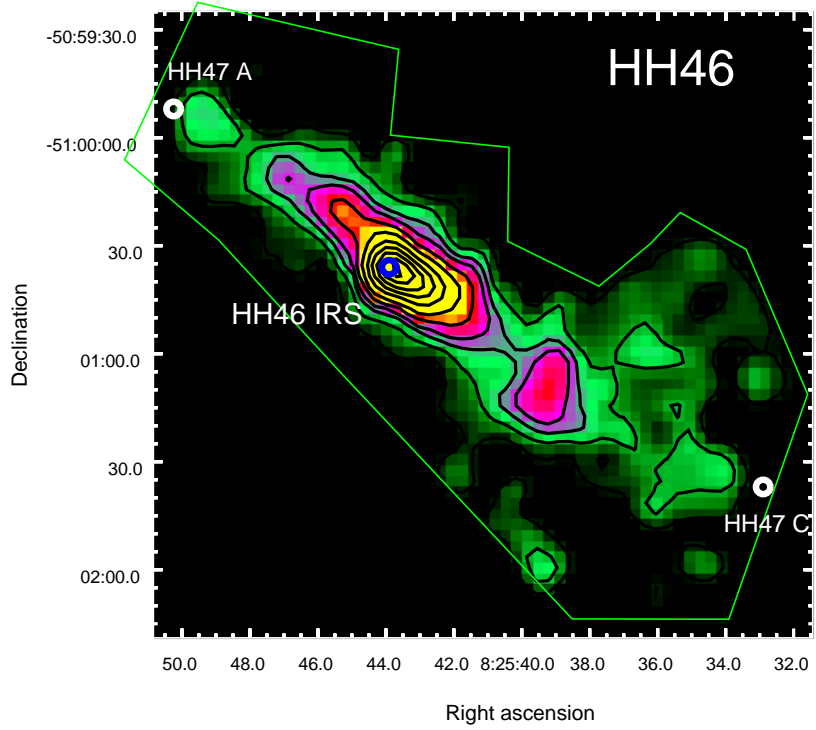}
\epsscale{0.7}
\plotone{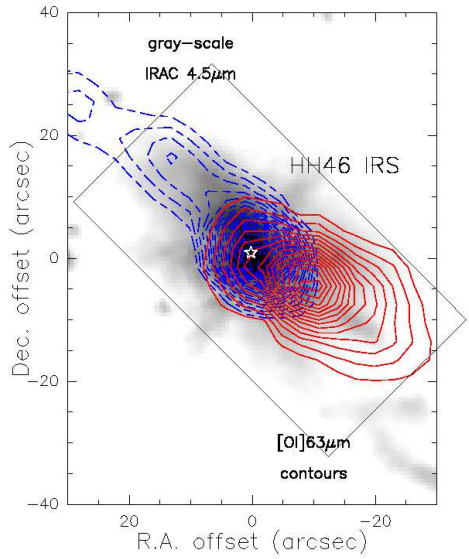}
\caption{As in Fig. 1 for HH46. In this case, however, the H$_2$O image is not available.
Blue- and red-shifted contours run from 2\,10$^{-5}$ to 2\,10$^{-4}$ \escs\,
in steps of 1\,10$^{-5}$ \escs\ . Velocity integration limits are -400 to 0 \kms\, and 0 to +400 \kms\, for the blue- and red-shifted emission, respectively.\label{fig1}}
\end{figure}

\begin{figure}
\epsscale{0.8}
\plotone{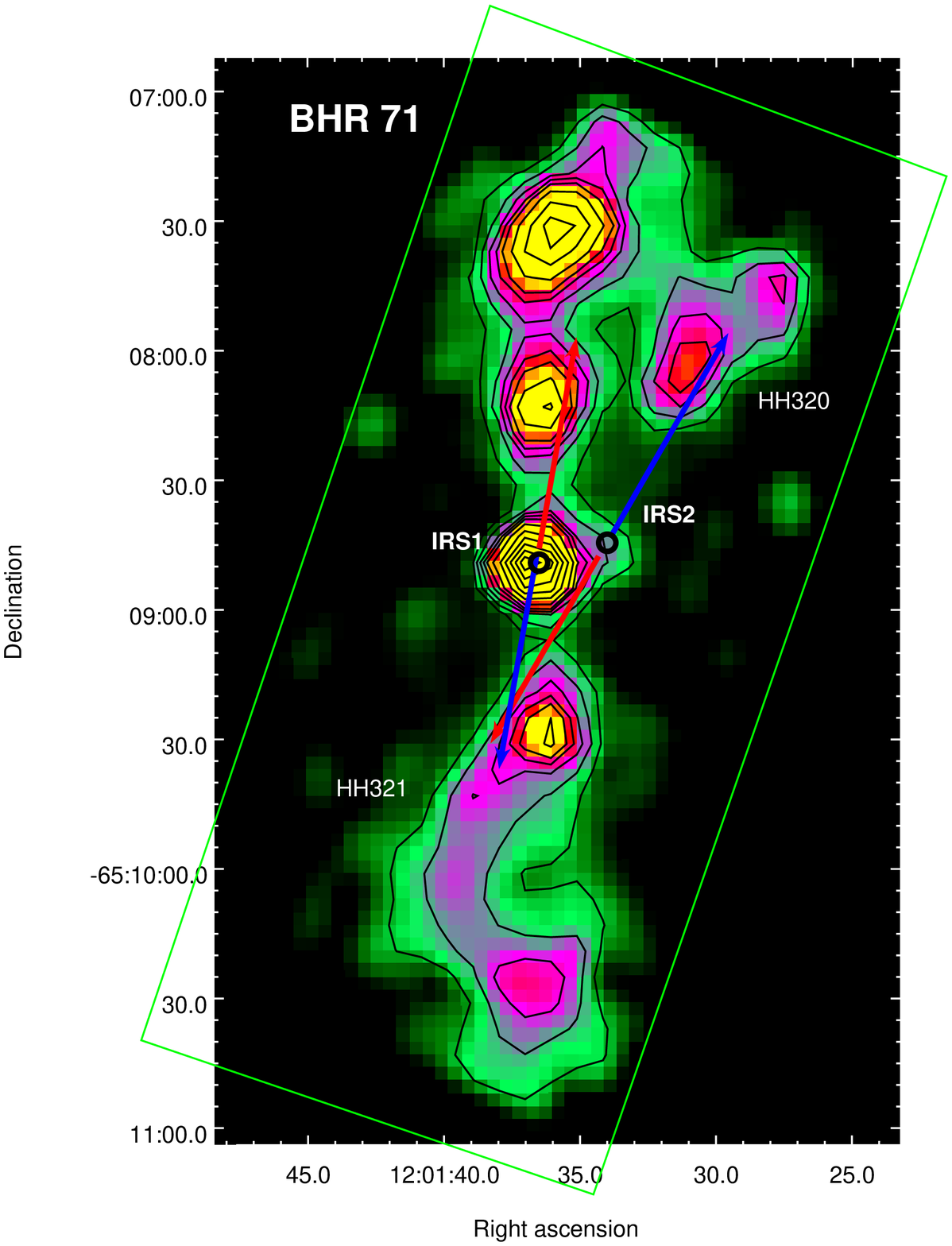}
\epsscale{0.7}
\plotone{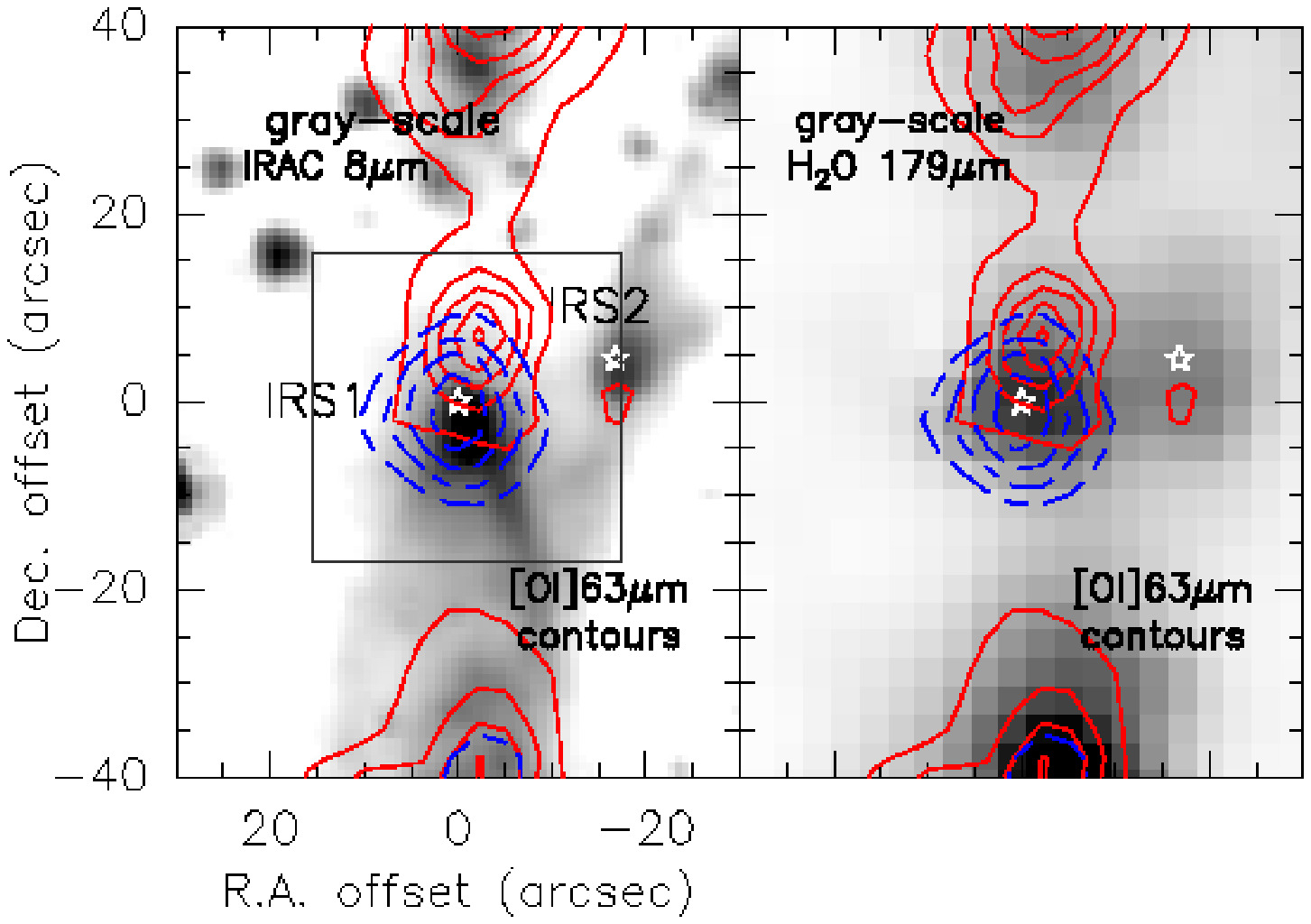}
\caption{As in Fig. 1 for BHR71. The arrows in the upper map indicate the directions
of the blue- and red-shifted lobes of the two outflows from the IRS1 and IRS2 sources.
In the bottom figure, red-shifted contours run from 1.5\,10$^{-5}$ to 7.5\,10$^{-5}$ \escs\,
in steps of 1.5\,10$^{-5}$ \escs\, while blue-shifted contours run from 1.0\,10$^{-4}$ to 4.0\,10$^{-4}$ \escs\,
 in steps of 1.0\,10$^{-4}$ \escs . Velocity integration limits are -200 to 0 \kms\, and 0 to +200 \kms\, for the blue- and red-shifted emission, respectively.\label{fig1}}
\end{figure}

\begin{figure}
\epsscale{1.}
\plotone{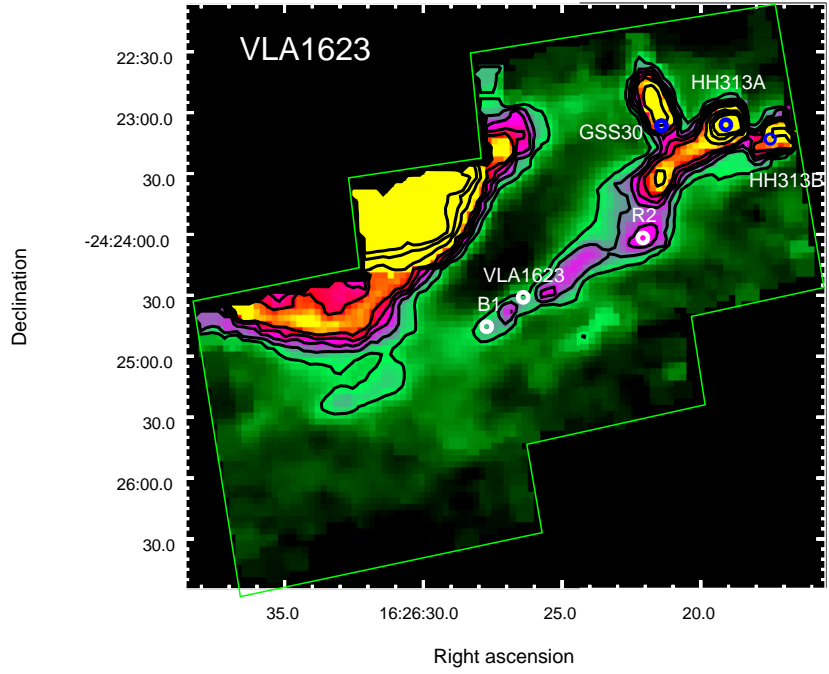}
\epsscale{0.7}
\plotone{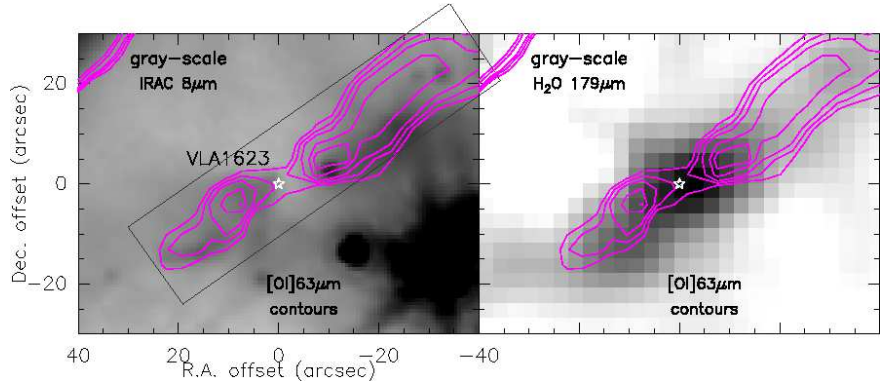}
\caption{As in Fig. 1 for VLA1623. Contours refer to
the total integrated line intensity and are drawn at the following values: 
1.2\,10$^{-4}$, 1.4\,10$^{-4}$, 1.8\,10$^{-4}$ and 2.0\,10$^{-4}$ \escs\,
\label{fig1}\label{fig1}}
\end{figure}

\begin{figure}
\epsscale{1.}
\plotone{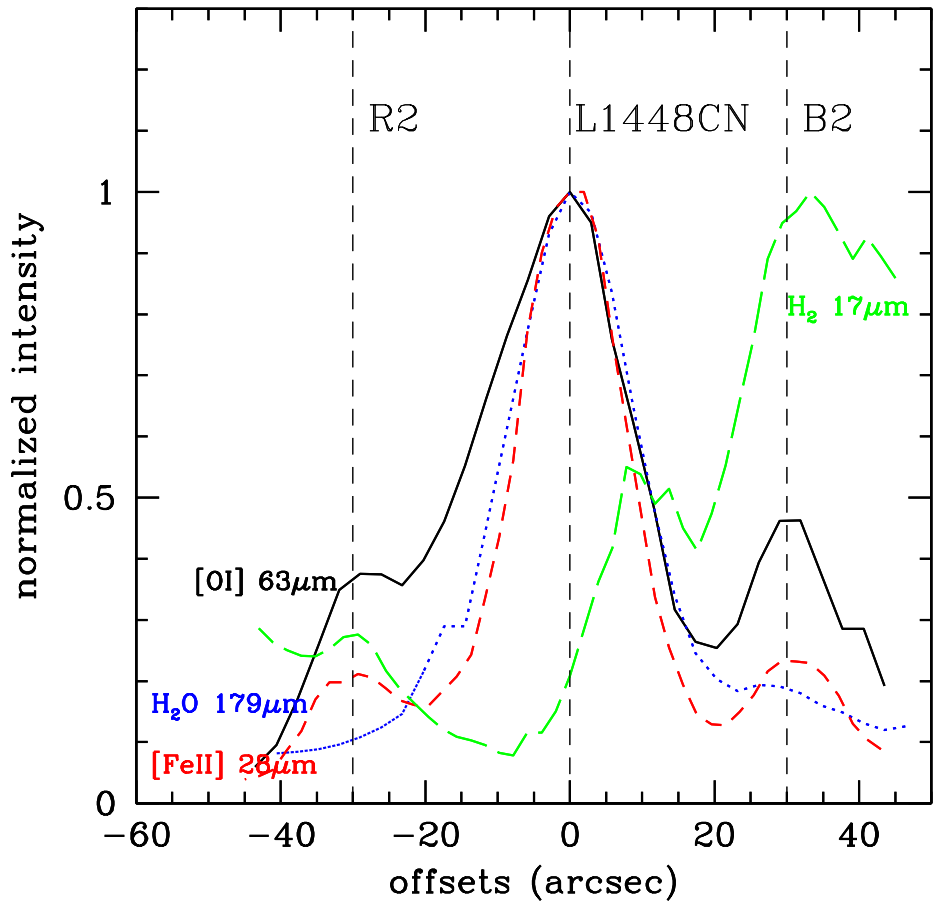}
\caption{
\Oxline\, line intensity cut of the L1448-C outflow along its axis (solid), compared with the intensity of the H$_2$O\,179 $\mu$m (dotted, from Nisini et al. 2013), the H$_2$\,17\,$\mu$m (long-dashed)
and [Fe II]\,28\,$\mu$m (dashed) emissions (from \textit{Spitzer}-IRS observations, Neufeld et al. 2010). The x-axis indicates
the offset, in arcsec, with respect to the L1448-CN source (at coordinates 03:25:38.8  +30:44:09.3). The offsets of the
two bright shock knots B2 and R2 are indicated with vertical lines 
\label{profile}}
\end{figure}

\begin{figure}
\epsscale{0.8}
\plottwo{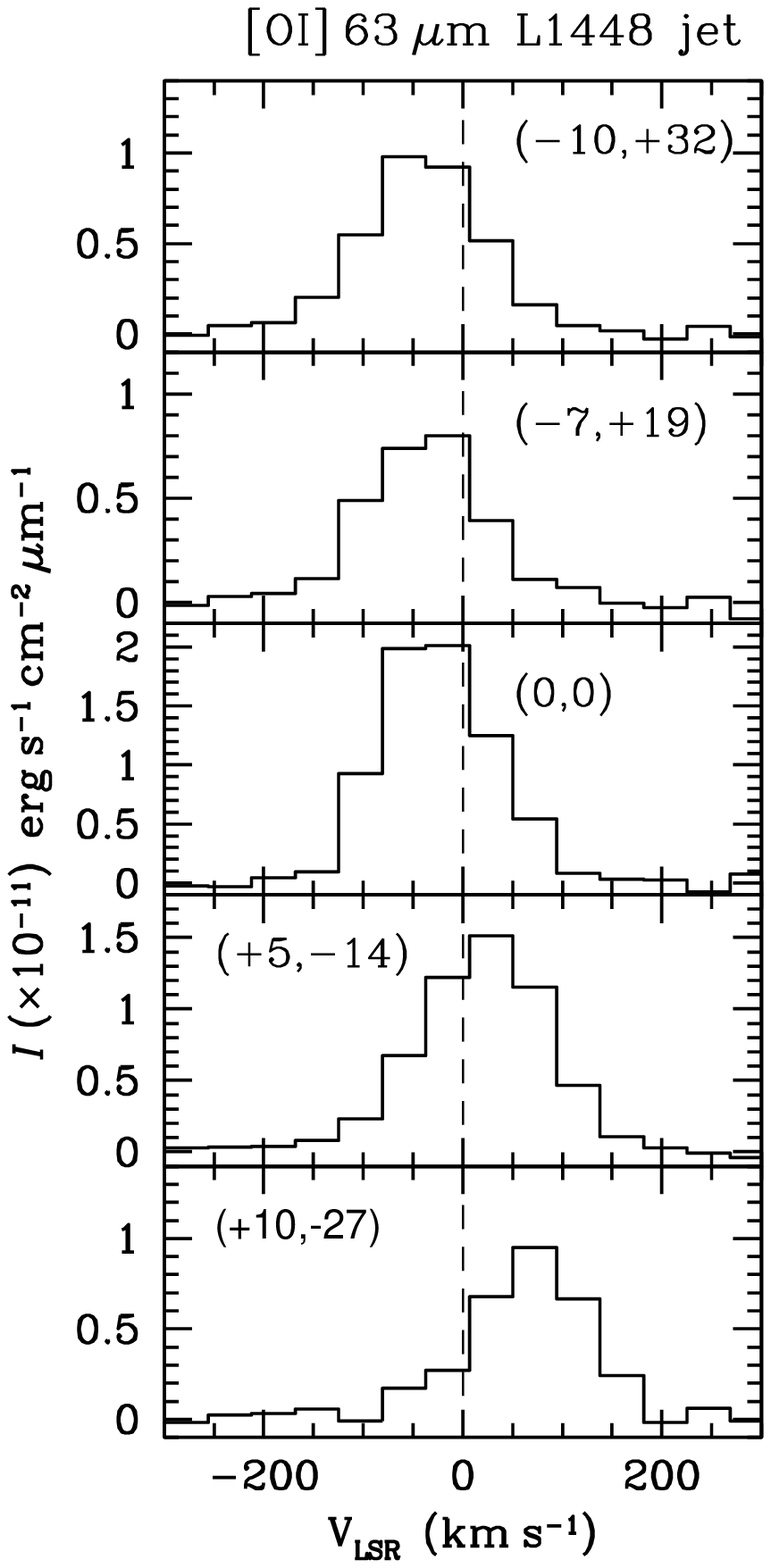}{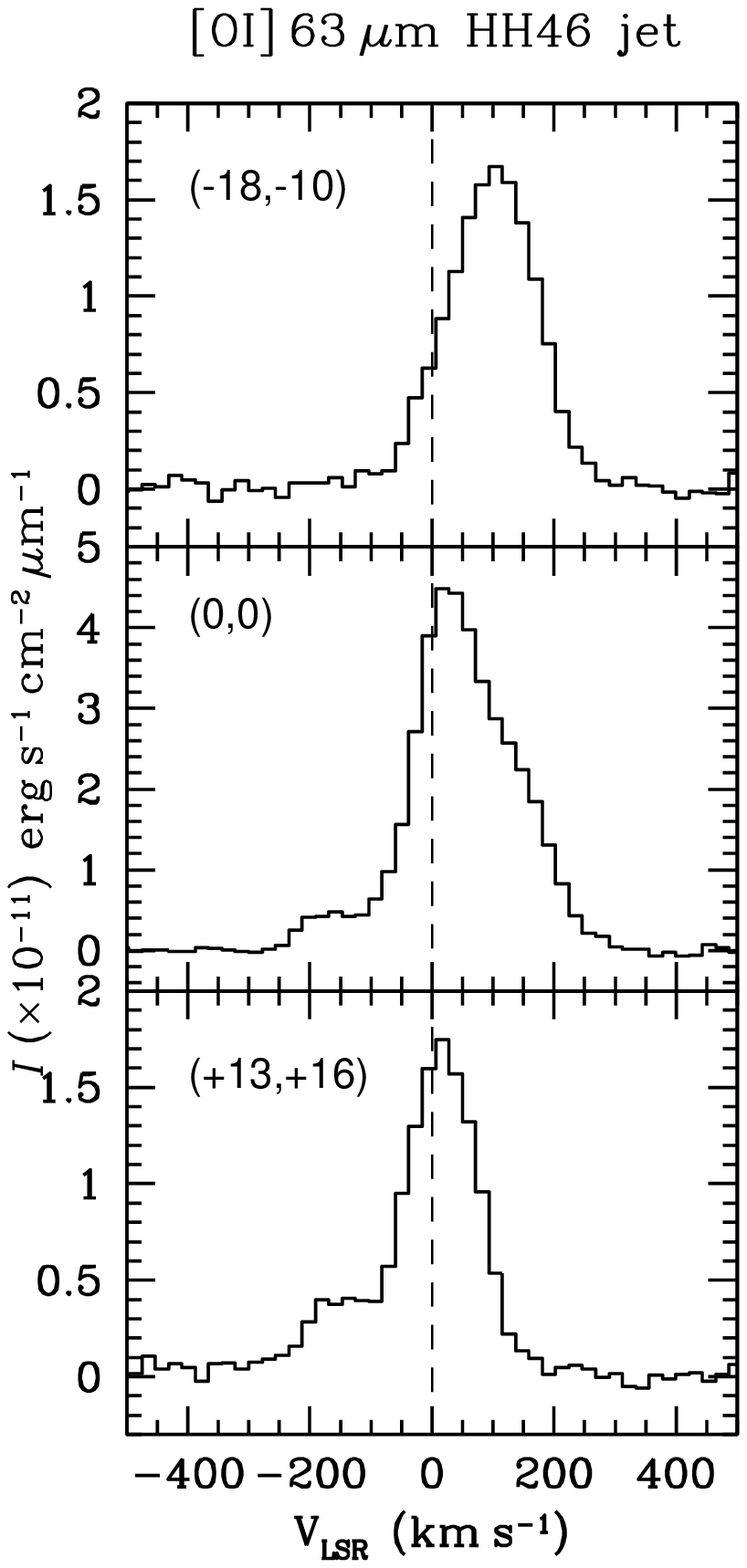}
\caption{Spectra of the \Oxline\, line of 15\arcsec $\times$ 15\arcsec\, rectangular regions
centered at the offsets with respect to the central source, indicated in each panel. Left for
L1448 and right for HH46. \label{fig5}}
\end{figure}

\begin{figure}
\epsscale{.60}
\plotone{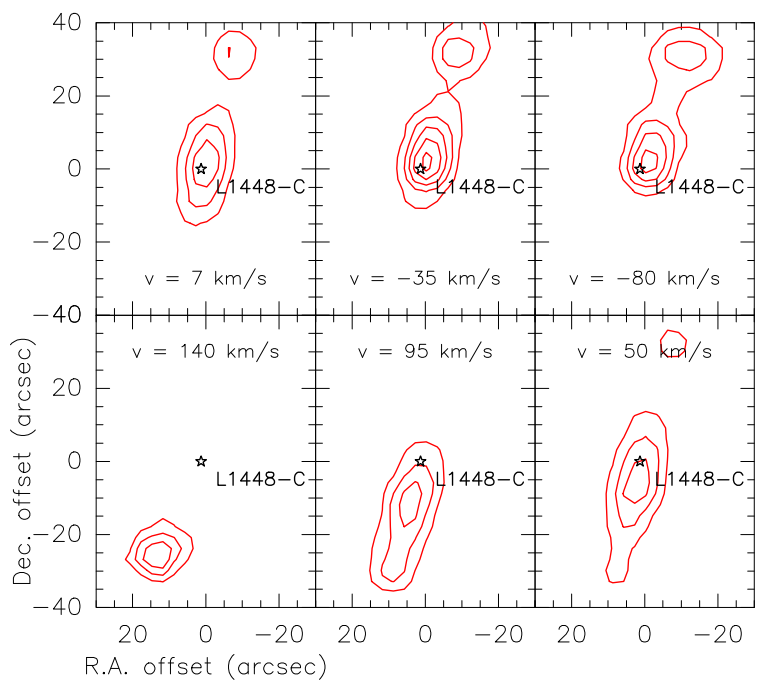}
\caption{Velocity channel maps of the L1448-C \Oxline\, emission. Each map represents a PACS spectral slice centered at the indicated velocity bin and covering half the nominal instrumental spectral resolution, i.e. about 44\kms.  \label{chan_1448}}
\end{figure}

\begin{figure}
\epsscale{.60}
\plotone{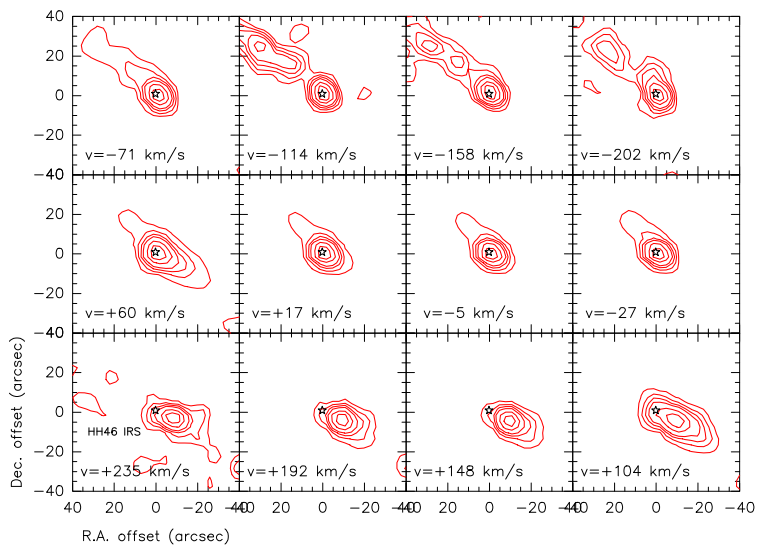}
\caption{As in Fig. \ref{chan_1448} for the HH46 flow.  \label{fig7}}
\end{figure}

\begin{figure}
\epsscale{1.2}
\plotone{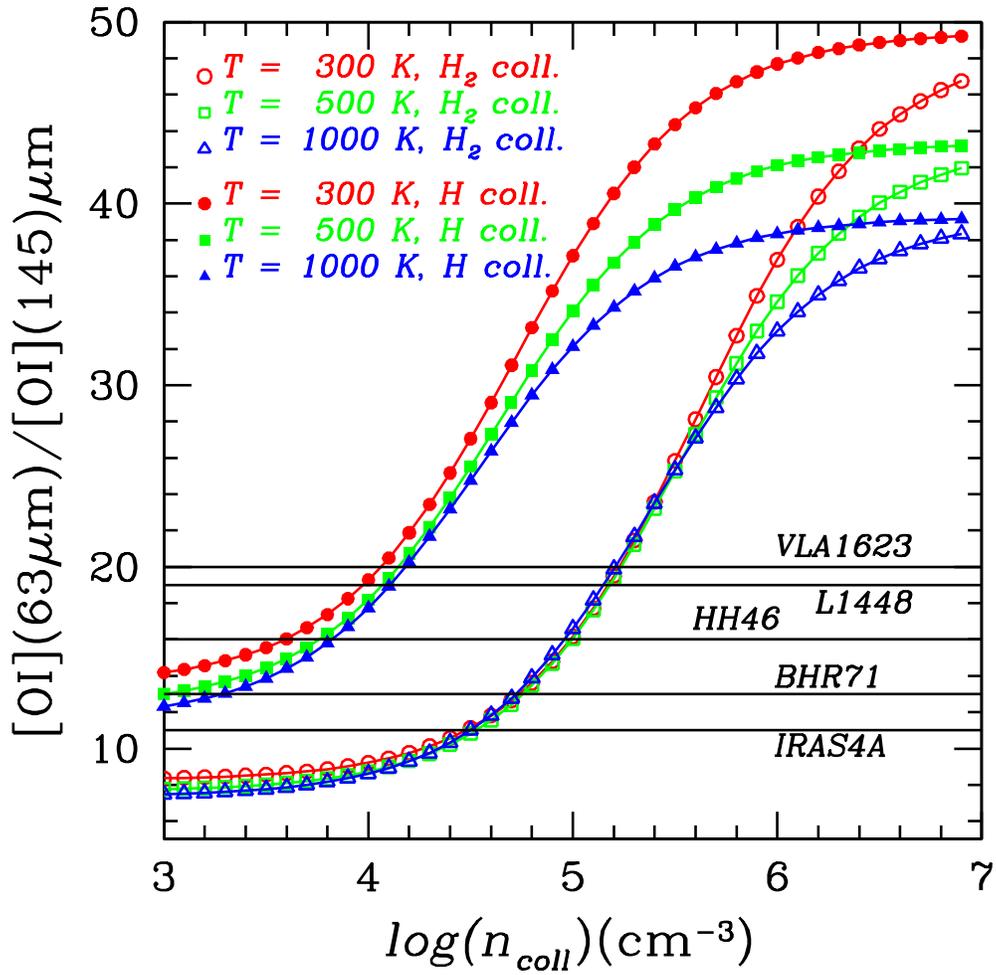}
\caption{[OI]\,63\um/145\um\, theoretical NLTE line intensity ratio plotted
as a function of density for three different temperature values . 
Open and filled symbols indicate the predictions assuming collisions with molecular
and neutral hydrogen, respectively. Horizontal lines indicate the values of the ratio
observed in the objects of our sample. Observational errors on the ratios are of the order of 20\%. 
Note that not knowing the composition
of the colliding particles introduces an uncertainty of about a factor of ten 
in the derived densities. \label{fig2}}
\end{figure}

\begin{figure}
\epsscale{1.2}
\plotone{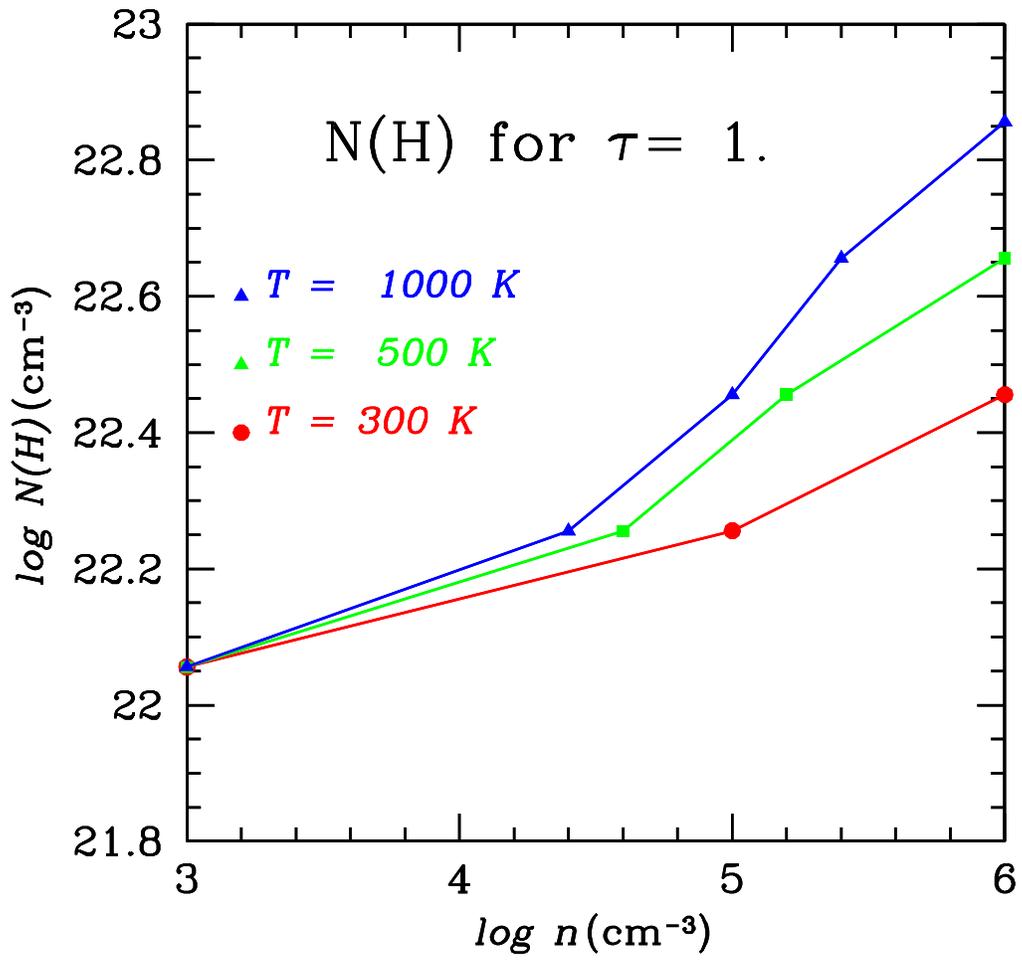}
\caption{Hydrogen column density versus volume density of atomic hydrogen, for unit optical depth in the \Oxline\, emission line. 
An oxygen abundance of 4.6\,10$^{-4}$ with respect to H
is assumed as well as a linewidth of 20\kms. \label{fig2}}
\end{figure}

\begin{figure}
\epsscale{1.3}
\plottwo{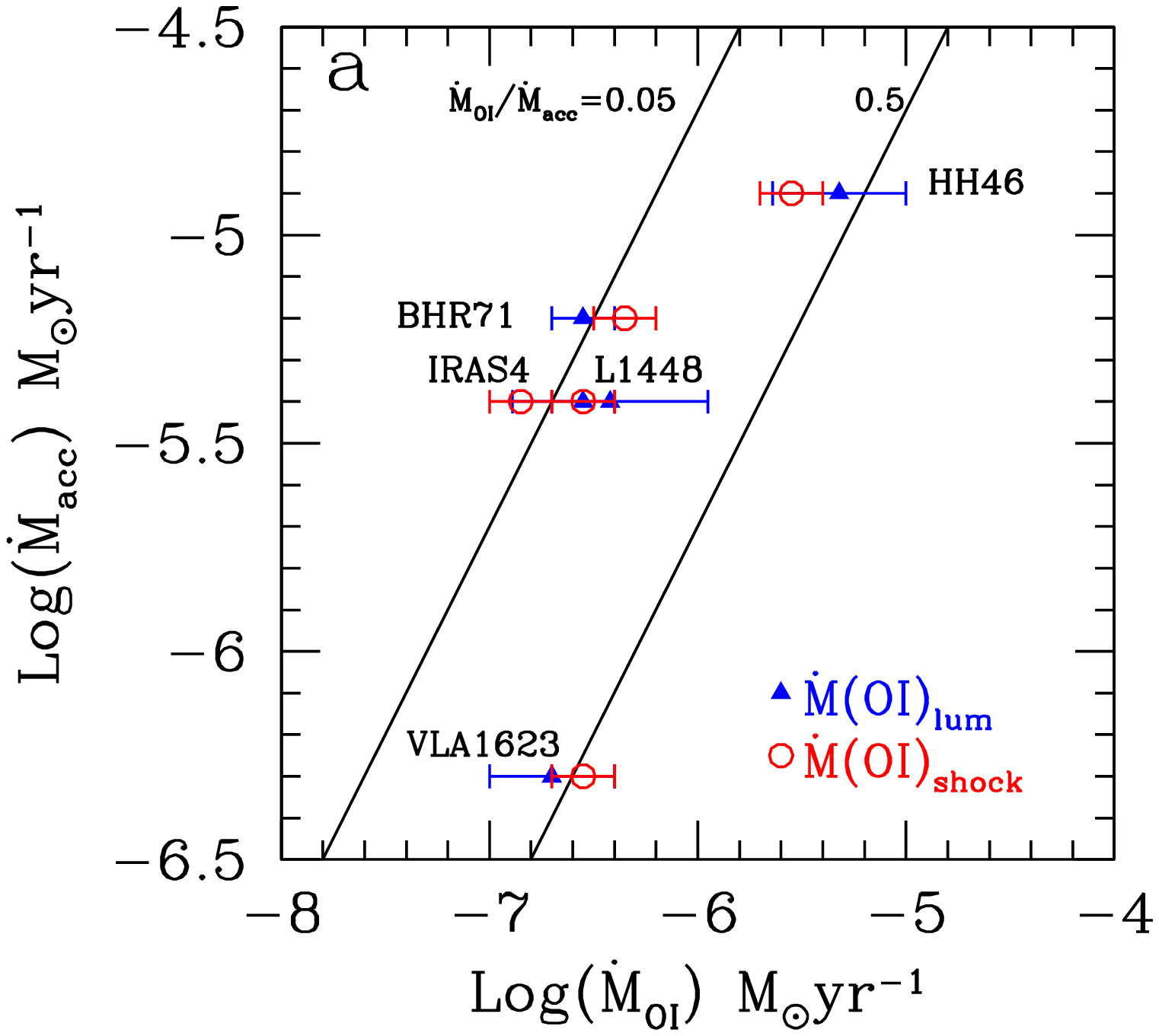}{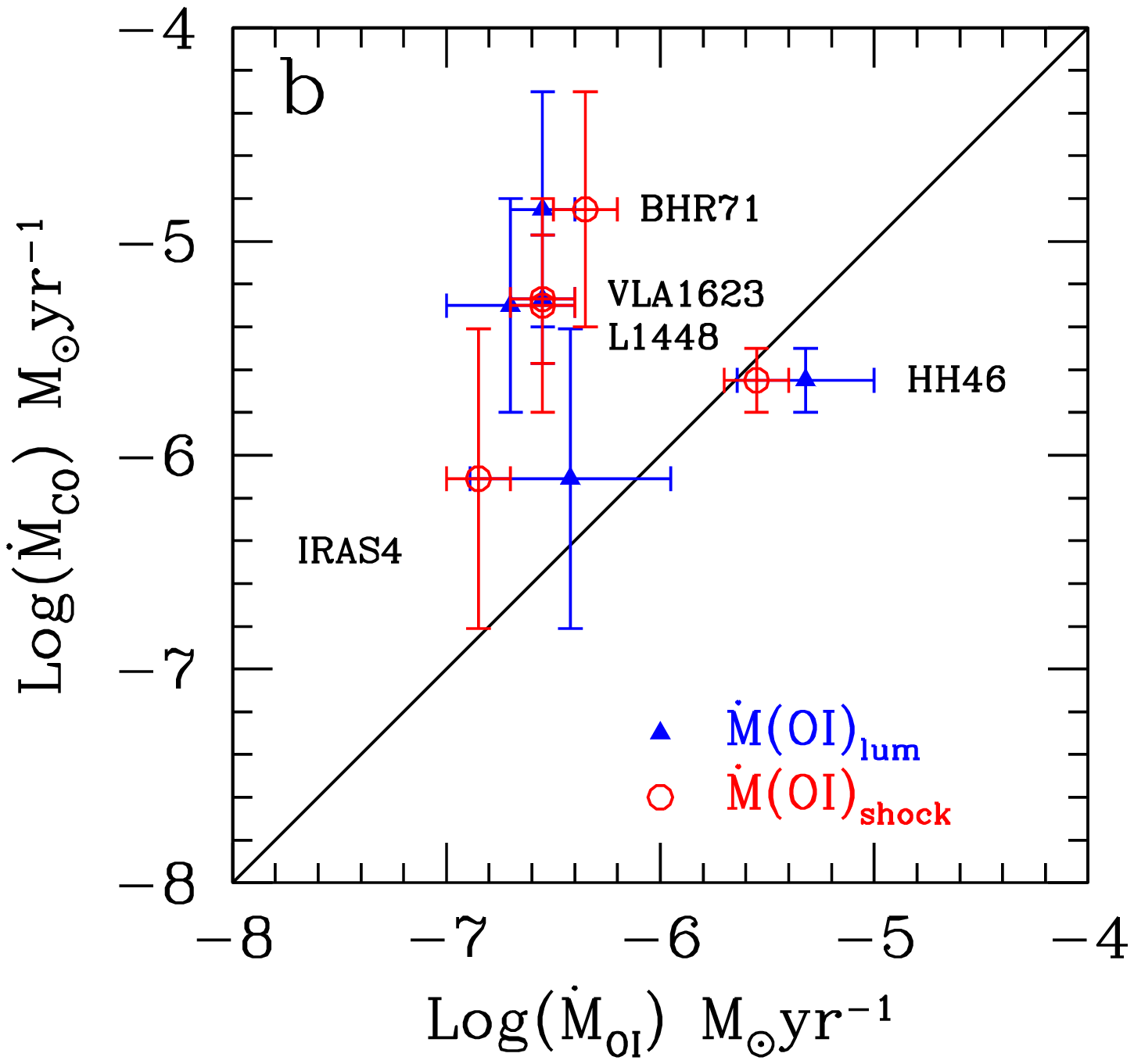}
\caption{Comparison between the [OI] atomic mass flux determinations and the protostellar 
mass accretion rate \textit{(a)}  and the wind mass
flux needed to sustain the CO outflow (\textit{(b)}, see details in Section 4). Blue triangles and red circles refer to
mean values of the determinations obtained from the \Oxline\, luminosity and from the Hollenbach  (1985)
formula that assumes that the 63\um\, line originates from the dissociative wind shock, respectively.\label{fig2}.
Error bars denote the range of the determinations.}
\end{figure}

\end{document}